\documentclass{emulateapj}
\usepackage{amsmath}
\usepackage{ulem}
\usepackage{CJK}
\shorttitle{Photosphere emission from GRBs} \shortauthors{Deng \& Zhang} \slugcomment{}
\begin{document}
\begin{CJK*}{UTF8}{gbsn}
\title{Low Energy Spectral Index and $E_{p}$ evolution of Quasi-thermal Photosphere Emission of Gamma-Ray Bursts}
\author{Wei Deng (邓巍), Bing Zhang (张冰)}
\affil{Department of Physics and Astronomy, University of Nevada Las Vegas, Las Vegas, NV 89154, USA\\deng@physics.unlv.edu, zhang@physics.unlv.edu}

\begin{abstract}
Recent observations by the Fermi satellite suggest that a photosphere emission component is contributing to the observed spectrum of many GRBs. One important question is whether the photosphere component can interpret the typical ``Band'' function of GRBs with a typical low energy photon spectral index  $\alpha \sim -1$. We perform a detailed study of the photosphere emission spectrum by progressively introducing several physical ingredients previously not fully incorporated, including the probability distribution of the location of a dynamically evolving photosphere, superposition of emission from an equal-arrival-time ``volume'' in a continuous wind, the evolution of optical depth of a wind with finite but evolving outer boundary, as well as the effect of different top-hat wind luminosity ($L_w$) profiles. By assuming a co-moving blackbody spectrum emerging from the photosphere, we find that for an outflow with a constant or increasing $L_w$, the low-energy spectrum below the peak energy ($E_{p}$), can be modified to $F_\nu \sim \nu^{1.5}$ ($\alpha \sim +0.5$). A softer ($-1<\alpha<+0.5$) or flat ($\alpha=-1$) spectrum can be obtained during the $L_w$ decreasing phase or high-latitude-emission-dominated phase. We also study the evolution of $E_{p}$ as a function of wind and photosphere luminosity in this photosphere model. An $E_p-L$ tracking pattern can be reproduced if a certain positive dependence between the dimensionless entropy $\eta$ and $L_w$ is introduced. However, the hard-to-soft evolution pattern cannot be reproduced unless a contrived condition is invoked. In order to interpret the Band spectrum, a more complicated photosphere model or a different energy dissipation and radiation mechanism are needed.
\end{abstract}

%\section{temp}
%1. using \\rm to write subscript

\section{Introduction\label{sec:intro}}

Despite more than 40 years of observations of gamma-ray bursts (GRBs), the radiation mechanism during the prompt emission phase is still being debated. Observationally the time-resolved spectra are usually characterized by a smoothly-joint broken power law, known as the ``Band'' function \citep{band93}. The typical value of the low-energy photon spectral index $\alpha$ (the one below the peak energy $E_p$) is around -1 \citep{preece00,zhang11,nava11}\footnote{Some GRBs show a harder $\alpha$ value in the time resolved spectra at the beginning or around the peak of the light curve \citep{ghirlanda02,ghirlanda03}. In some other cases, a blackbody component can play a dominant role \citep{ryde10,zhang11,ghirlanda13}.}. This value is not straightwardly predicted in available models. In general there are two broad categories of GRB prompt emission models, one invoking a Comptonized quasi-thermal emission from the photosphere of the outflow \citep{thompson94,ghisellini99,meszarosrees00,meszaros02,rees05,peer06,thompson07,peer08,giannios08,beloborodov10,lazzati10,ioka10,toma11,peer11,mizuta11,lundman13,lazzati13,ruffini13}, the other invoking a non-thermal (synchrotron or synchrotron-self-Compton) mechanism in the optically thin region \citep{meszaros94,tavani96,daigne98,lloyd00,zhangmeszaros02c,kumarpanaitescu08,zhangpeer09,wang09,kumarnarayan09,daigne11,zhangyan11,zhang12,veres12b,burgess11,uhm13}. Recent observations with Fermi \citep{ryde10,zhang11,guiriec11,axelsson12,guiriec13} confirmed the pre-Fermi suggestion \citep{ryde05,ryde09} that the observed GRB spectra sometimes include superposition of a quasi-thermal component with a non-thermal component\footnote{Observational constraints demand that non-thermal component cannot be a single power law extending to much lower energies \citep{ghirlanda07}.}, suggesting that the photosphere emission indeed contributes to the observed GRB spectra of at least some GRBs. The question is how much the photosphere emission contributes to the observed spectra. There are two different opinions. According to the first opinion, only the quasi-thermal component identified in some GRBs is of the photosphere origin \citep[e.g.][]{ryde10,zhang11,guiriec11,axelsson12,guiriec13}, while the main ``Band'' component is non-thermal emission from an optically thin region. The second, more optimistic opinion suggests that the main Band component is of a photosphere origin, and the observed $E_p$ is defined by the photosphere temperature \citep[e.g.][]{rees05,thompson07,beloborodov10,lazzati10,peer11,lundman13,lazzati13}.

Within this second scenario, one needs to broaden the quasi-thermal spectrum to a mimic the observed ``Band'' function. The high-energy photon index can be easily interpreted by introducing dissipation of energy near photosphere so that non-thermal electrons are accelerated to upscatter the seed quasi-thermal photons \citep[e.g.][]{lazzati10}. However, the main difficulty is to account for the observed low-energy photon index $\alpha \sim -1$, since the predicted value is much harder than this value \citep[e.g.][]{beloborodov10}. It has been speculated that geometric smearing and temporal smearing may soften the spectrum to make $\alpha$ closer to -1, but no thorough study has been carried out.

Another interesting observational feature is the $E_p$ evolution in GRBs. Observtionally two patterns are identified: hard-to-soft evolution or $E_p$-intensity tracking \citep{liang96,ford95,ghirlanda10,ghirlanda11a,ghirlanda11b,lu10,lu12} across a broad GRB pulse \citep[or the ``slow'' variability component,][]{gao12}. Various data analyses suggest that both patterns co-exist, sometimes in a same burst. For short GRBs, the tracking behavior is most common, while for long GRBs, especially for the first broad pulse, the hard-to-soft evolution is relatively more common \citep[e.g.][]{preece14}. Simulations show that in long GRBs, tracking pulses after the first pulse could be due to a superposition of multiple hard-to-soft pulses \citep{lu12}, and it has been argued that the hard-to-soft evolution pattern may be ubiquitous among long GRBs \citep{hakkila11,hakkila14}. It is of great interest to see what $E_p$ evolution pattern the photosphere model predicts.

In this paper, we investigate the predictions of $\alpha$ values and $E_p$ evolutions within the simplest photosphere model (a co-moving blackbody spectrum and a uniform jet) by fully treating the geometrical and temporal smearing effects. In Sect.2, we describe our methodology, in particular, improvements upon previous work. We then present calculations of photosphere spectra and $E_p$ evolution patterns in progressively more complicated models in Sect.3. The conclusions are drawns in Sect.4.

\section{Methodology}

\subsection{Previous work on probability photosphere model}

Photosphere may be roughly defined as a radius at which the Thomson scattering optical depth for a photon is $\tau = 1$. From the microscopic view, an individual photon in the outflow can be in principle last scattered by an electron at any position with a certain probability. One should introduce a probability function to describe photosphere emission \citep{peer08,peer11,beloborodov11}. In general, for a group of photons emerging from deep inside the photosphere, they can undergo last scattering at any location ($r,\Omega$) inside the outflow, where $r$ is the radius from the explosion center, and $\Omega (\theta, \phi)$ defines the angular coordinates. A probability function $P(r,\Omega)$ is introduced to describe the chance of last scattering at any location.

\begin{figure}[ht]
\includegraphics[angle=0, width=3.5in]{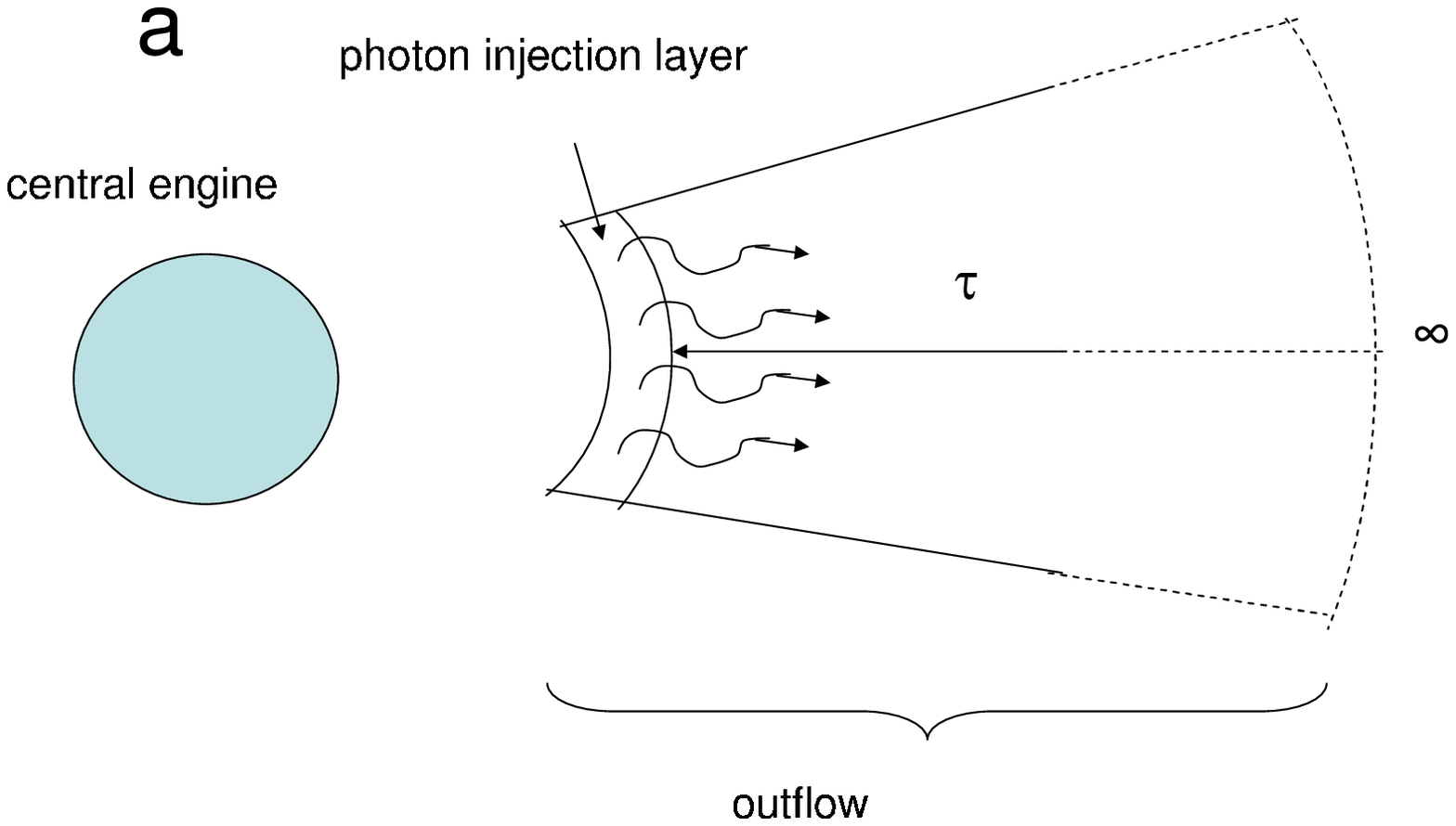}'
\includegraphics[angle=0, width=4.0in]{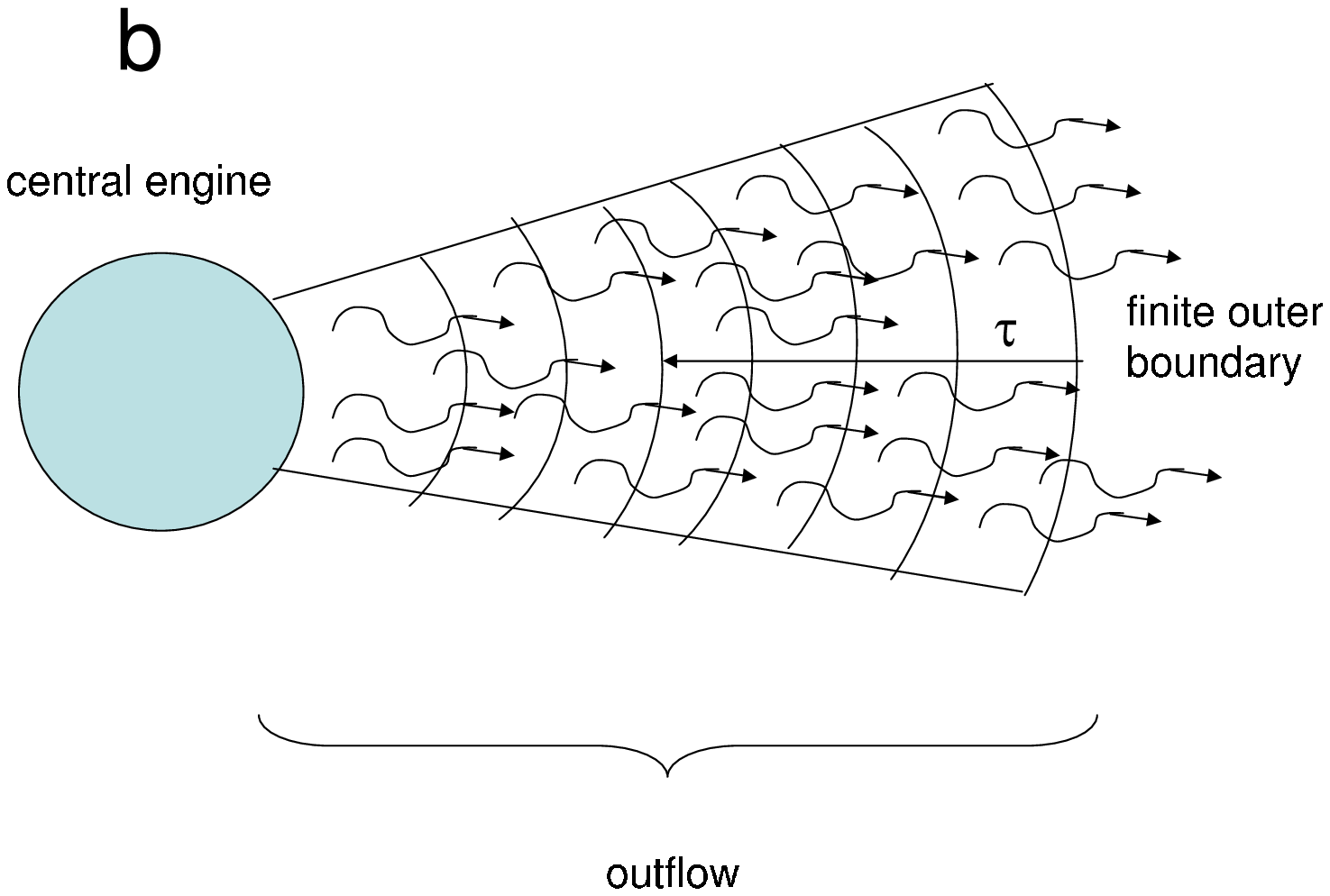}
\caption{The carton picture for the model of \cite{peer08,peer11} (a) as compared with ours (b).}
\label{model_peer}
\end{figure}

So far, a most detailed treatment of the probability function was presented in \cite{peer08}, \cite{peer11} and \cite{beloborodov11}. By making some simplifications in modeling, their model caught the essence of the photospheric physics. The basic physical picture of \cite{peer11} can be summarized as follows (see the cartoon picture Figure \ref{model_peer}a): The central engine remains active for a while and forms a conical outflow. After some time the central engine is shut down. They assumed that the wind luminosity $L_w$, the initial mass loss rate $\dot M$, and hence, the initial bulk lorentz factor $\Gamma$, are all constant. This means that a continuous wind with a time-independent density profile is established. For simplicity, they also assumed that such a wind extends to an infinite distance. As a result, the optical depth could be directly calculated by an analytical formula, which is time-independent \citep{peer08,peer11}. Another simplification of \cite{peer11} is that they only considered an instantaneous deposition of photons in one thin layer at the center of an expanding outflow plasma. In reality, photons are continuously deposited into a series of layers ejected by a long-lasting central engine. This would make calculations more complicated. A third simplification in their analytical approach is that they assumed a mono-energetic photon field, while in reality the photons have a distribution (e.g. blackbody in the co-moving frame). This assumption was removed in their numerical approach through Monte Carlo simulations. In any case, an analytical approach to handle the blackbody spectrum is welcome.

\subsection{Improvements}\label{sec:improvements}

In order to give a more precise treatment on more realistic situations, in this paper we make several critical improvements on the previous work to establish a more  sophisticated photosphere model to study the instantaneous and time-integrated spectra and $E_{p}$ evolution by allowing a time-varying central engine wind, and hence, the optical depth. Our improvements on the previous work include the following: 1. We introduce a blackbody distribution of the photons in the comoving frame, to replace the mono-energetic photon distribution in the analytical treatment of \cite{peer08} and \cite{peer11}; 2. For a wind lasting for a certain duration, at any instant, photons from different layers and latitudes, which were emitted at different epochs, are received. Instead of the traditional ``equal-arrival-time-surface'' effect commonly considered in an afterglow model, one needs to consider ``equal-arrival-time-volume'' for a more precise treatment of the photosphere emission. We separate the outflow into many thin layers, follow the last scattering of electrons from each layer individually, and calculate the sum of flux from all the layers from different latitudes and emission times, but at a same observer time; 3. We introduce a finite, dynamically evolving front of the outflow based on the assumed injection history of the outflow. This would affect the calculation of the Thomson scattering optical depth of the photons, which leads to a more precise derivation of the photosphere radius in different directions. The effect is especially important in the early phase of the outflow; 4. Since GRB lightcurves show an erratic behavior, we introduce time-dependent wind luminosity and baryon loading to allow a more realistic treatment of time-variable photosphere emission.

A cartoon picture of our improved model is shown is Fig.\ref{model_peer}b, as compared with Fig.\ref{model_peer}a. Due to the finite central engine activity time, the outflow has a finite thickness. We separate the outflow into many thin layers, each characterized by an initial wind luminosity $L_w$ and an initial mass loading rate $\dot{M}$. To treat the equal-arrival-time volume effect, we consider two levels of integration. The first level handles the equal-arrival-time surface of each layer: photons emitted from different latitudes at different times arrive the observer at the same time; The second level handles superposition of emission from different layers: high-latitude emission from an earlier layer would arrive at the same time with low-latitude emission from a later layer.

In order to analytically treat the problem, some assumptions still need to be made. One important parameter is the Lorentz factor $\Gamma$ of each layer. According to the standard fireball model \citep{meszaros93,piran93,kobayashi99}, $\Gamma$ should initially increase with radius before coasting to a certain value. The coasting $\Gamma$ depends on a comparison between the dimensionless entropy of the fireball $\eta$ and a critical value $\eta_c$ \citep{meszarosrees00}. If $\eta < \eta_c$, the photosphere radius $r_{ph}$ is above the saturation radius $r_s = \eta r_0$ (where $r_0$ is the radius of the central engine), and the Lorentz factor coasts to a constant value $\Gamma_0 = \eta$. For $\eta > \eta_c$, the photosphere occurs during the fireball acceleration phase ($r_{ph} < r_s$). The Lorentz factor $\Gamma$ would vary significantly across $r_{ph}$, which is difficult to handle analytically. In our treatment, we have made the assumption of a constant $\Gamma_0$ throughout the shell evolution (i.e. the acceleration phase is neglected).  In our model, for a certain layer although all the positions have a certain probability to release the last-scattered photon, the maximum probability is concentrated around the photosphere radius $r_{ph}$. So our calculation of the emerging photosphere spectrum based on the assumption of a constant $\Gamma_0$ is accurate enough as long as $r_{ph} \gg r_s$, but would become progressively inaccurate as $r_{ph}$ becomes progressively smaller. In our calculations, this constraint is considered in care, so that the presented spectra are all in the regime where this assumption is valid.

\section{Models \& Results \label{sec:model}}

In this section, we calculate the photosphere spectra by progressively including the four new physical ingredients as discussed above. In \S\ref{sec:bb} we introduce a blackbody spectrum in the comoving frame, while still keeping the basic assumptions of \cite{peer11} such as a constant wind luminosity, infinity outer boundary and single layer emission. In \S\ref{sec:wind} we introduce multiple layers to study the superposed spectrum. In \S\ref{sec:variable_boundary}, keeping the wind luminosity constant, we introduce a time-evolving outer boundary of the outflow. In \S\ref{sec:variable_luminosity} we introduce a time-dependent wind luminosity and time-evolving outer boundary of the outflow at the same time. A discussion on $E_p$ evolution patterns are presented in \S\ref{sec:Ep-patterns}.

\subsection{Impulsive injection, outer boundary at infinity, blackbody distribution of photons}\label{sec:bb}

In this section we still keep most simplifications made by \cite{peer11}, including instantaneous injection of photons and an outer boundary at infinity. One modification is that we relax the $\delta$-function assumption of photon energy \citep{peer08,peer11}, but introduce a more realistic co-moving blackbody spectrum.

We start with the formula of \cite{peer11} that calculates the specific flux:
\begin{eqnarray}
F_\nu (\nu,t) &  =& {N_0 \over 4 \pi d_{\rm L}^2} \int \int P(r,\theta) k_{\rm B} T(r,u)
\nonumber \\
&\times &\delta \left(t - {r u \over \beta c} \right) \delta \left(\nu - {k_{\rm B} T'(r) \over h \Gamma u} \right) du dr
\label{eq:peer}
\end{eqnarray}
where $N_0$ is the impulsively injected number of photons at $t=0$, $P(r,\theta)$ is the probability density function for the last scattering to occur at the radius $r$ and to an angle $\theta$, respectively, and the parameter
\begin{equation}
u=1-\beta \cos\theta
\end{equation}
contains the angular information.

For the  probability density function $P(r,\theta)$, \cite{peer11} used two independent functions $P(r)$ (for the radial dimension) and $P(u)$ (for the angular dimension) to decompose $P(r,\theta)$ into $P(r,\theta)=P(r) \cdot P(r)$, where
\begin{equation}
P(r)={r_{\rm ph} \over r^2} e^{-{r_{\rm ph} \over r}}
\label{Pr}
\end{equation}
where $r_{\rm ph}$ is the classical photosphere radius ($\tau=1$) at $\theta=0$, and
\begin{equation}
P(u)={1 \over 2 \Gamma^2 \beta u^2}.
\label{Pu}
\end{equation}
On the other hand, the simulation result in \cite{peer11} indicated that $P(r)$ depends on angle and $P(u)$ also depends on radius. \cite{beloborodov11} later introduced a two dimensional $P(r,\theta)$ function based on the assumption of outer boundary at infinity:
\begin{eqnarray}
\frac{dP}{dr d\mu}&=&{\cal D}^2\frac{r_{\rm ph}}{4r^2}
\left\{\frac{3}{2}+\frac{1}{\pi}\arctan\left[\frac{1}{3}
\left(\frac{r_{\rm ph}}{r}-\frac{r}{r_{\rm ph}}\right)\right]\right\}\nonumber \\
&&\exp\left[-\frac{r_{\rm ph}}{6r}
\left(3+\frac{1-\mu'}{1+\mu'}\right)\right],
\label{eq:Pru_Bel}
\end{eqnarray}
where $\mu'=cos\theta'$ is in the outflow comoving frame, $\mu=cos\theta$ is in the observer frame, and the Doppler factor is
\begin{equation}
{\cal D} = [\Gamma(1-\beta\cos\theta)]^{-1} = (\Gamma u)^{-1}.
\end{equation}
In this section and \S\ref{sec:wind}, we will use Eq.\ref{eq:Pru_Bel} as $P(r,\theta)$ in Eq.\ref{eq:peer}.

The last $\delta$ function in Eq.\ref{eq:peer} made the mono-energetic simplification for the photosphere photons. We consider a blackbody distribution of the photon energy and replace Eq.\ref{eq:peer} by a new equation
\begin{eqnarray}
F_\nu (\nu,t) & =& {N_0 \over 4 \pi d_{\rm L}^2} \int \int P(r,\mu) P(\nu,T)h\nu \nonumber \\
&\times &\delta \left(t - {r u \over \beta c} \right)d\mu dr,
\label{eq:Fnu1}
\end{eqnarray}
where we have defined a new parameter
\begin{equation}
P(\nu,T) = { n_\gamma(\nu,T) \over \int_{0}^\infty n_\gamma(\nu,T) d\nu}
 =  \frac{n_\gamma (\nu, T) (hc)^3}{16 \pi \zeta(3) (kT)^3},
\end{equation}
which is the probability function of a photon with frequence $\nu$ in a Plank distribution with a temperature $T$ at the coordinate $(r,\theta)$ as observed by an observer located at $\theta=0$, and
\begin{equation}
n_\gamma(\nu,T) = \frac{8\pi \nu^2}{c^3}\frac{1}{\exp(h\nu/kT) -1}
\end{equation}
is the specific photon number density at frequency $\nu$ for an observed temperature $T$. Here the mathematical relation
\begin{equation}
\int_{0}^\infty {x^2 dx \over e^{\rm x}-1}=2\zeta(3)=2 \times 1.202...
\end{equation}
has been applied when calculating the integration $\int_0^\infty n_\gamma(\nu,T) d\nu$. Notice that here we have also adopted a blackbody function for the observed spectrum at any spatial point $(r, \theta)$. This is justified, since for any spatial point, the spectral shape is not modified from that in the co-moving frame, except that the entire spectrum is Doppler boosted by the local Doppler factor \citep{lisari08}. The global observed spectrum deviates from a blackbody due to different Doppler factors at different points. This effect is fully incorporated in our calculations.

The observer frame temperature at point $(r, \theta)$
\begin{equation}
T(r, \theta) = {\cal D} T'(r)
\end{equation}
depends on the angle through the Doppler factor and on the co-moving temperature. The co-moving frame temperature $T'(r)$ is more intrinsic, which depends on the radius $r$ only. Its expression depends on the radius range \citep{meszarosrees00}, and can be calculated using the on-axis observed temperature divided by the on-axis Dopper factor ${\cal D}(\theta=0) = 2\Gamma$, i.e.
\begin{eqnarray}
T'(r) & = & {T(r,\theta=0) / (2\Gamma)} \nonumber \\
& = & \left \{
 \begin{array}{lll}
 {T_0 / (2 \Gamma)},& r<r_{\rm s}<r_{\rm ph},
\\ \\
  {T_0{(r/r_{\rm s})}^{-2/3} / (2 \Gamma)},& r_{\rm
s}<r<r_{\rm ph},
 \\ \\
  {T_0{(r_{\rm ph}/r_{\rm s})}^{-2/3} / (2 \Gamma)}, & r_{\rm
s}<r_{\rm ph}<r,
   \end{array}
   \right.
\label{eq:comoving_T}
\end{eqnarray}
where
\begin{equation}
T_0 = \left(\frac{L_0}{4 \pi {r_0}^2 c a}\right)^{1/4}
\label{eq:T}
\end{equation}
is the temperature at the central engine, $L_0$ is the initial luminosity deposited at the central engine, and $r_0 = 10^7 r_{0,7}$ is the central engine radius.  Notice that in order to satisfy our constant $\Gamma$ assumption (\S\ref{sec:improvements}), we have limited our study in the regime $r_s < r_{ph}$. Also for an easy treatment, in the above analytical model $T'(r)$ is taken as a broken power law function of $r$, whereas in reality it is a smoothly connected broken power law \citep{peer08}.

Since we mostly care about the shape of the spectrum, the normalization parameter $N_0$ is approximately taken as
\begin{equation}
N_0={E_0 \over k T_0},
\label{eq:N_0}
\end{equation}
which denotes the rough total number of photons released at the central engine (assuming mono-energetic photon energy).  By doing so, we have assumed that no additional emission or absorption processes occur as the photon-mediated outflow travels from the central engine to the photosphere, and that photons only undergo Thomson scattering with the total number conserved.

\begin{figure}[ht]
\includegraphics[angle=0,scale=0.45]{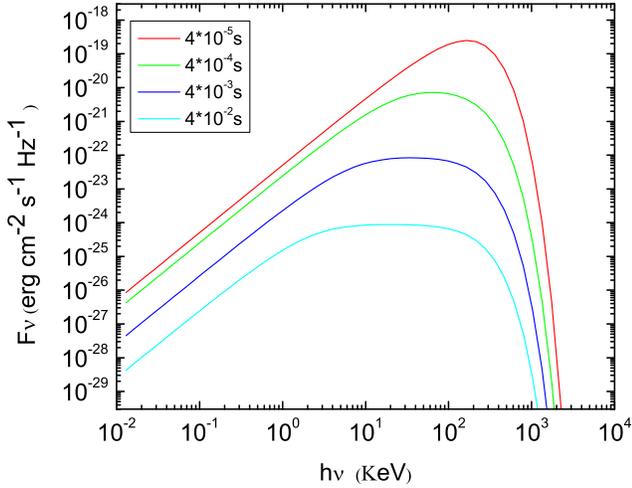}
\caption{Instantaneous photosphere spectra for a fireball with impulsive injection of energy. The impulsively injected total energy is $E_0=10^{52}$ erg, the fireball temperature is calculated by taking $L_0 = 10^{52}~{\rm erg~s^{-1}}$, central engine radius $r_0 = 10^7$ cm, dimensionless entropy $\eta=\Gamma_0= 300$, and luminosity distance $d_{\rm L} = 2\times 10^{28}$ cm. Different colors represent different observational times. The spectra become progressively high-latitude dominated.}
\label{fig:bb}
\end{figure}

We integrate Eq. \ref{eq:Fnu1} to calculate the instantaneous spectra at different times. The angle $\theta$ is integrated from 0 to $\pi/2$, which is wide enough to catch the relativistically beamed emission. The range of $r$ is defined by the equal-arrival-time equation
\begin{equation}
t = \frac{ru}{\beta c}.
\end{equation}
The results are presented in Fig.\ref{fig:bb}. For an impulsive fireball as studied in this subsection, we calculate $T_0$ by taking $L_0 = 10^{52}~{\rm erg~s^{-1}}$, while adopt the impulsively injected total wind energy $E_0 = 10^{52}$ erg. This is to keep consistency with the continuous-wind calculations in the later subsections. Other parameters are adopted with the following values: dimensionless entropy $\eta=\Gamma=300$, luminosity distance of the GRB $d_{\rm L}=2\times 10^{28}$ cm $(z\sim 1)$, and the inner boundary of integration set to $r_0=10^7$ cm.

From Fig.\ref{fig:bb}, we can see that under the assumptions adopted in this subsection, the on-axis, instantaneous photosphere spectrum evolves from a pure blackbody (early on) to a gradually flattened shape as the high-latitude emission becomes progressively dominant. Compared with the analytical results of \cite{peer11}, our results show an exponential tail of blackbody emission instead of the flat spectrum $F_\nu \propto \nu^0$ extending all the way to high energies. Our results are however generally consistent with the numerical results of \cite{peer11}.

\subsection{Continuous wind with a constant wind luminosity and Lorentz factor}\label{sec:wind}

The next step is to study the observed instantaneous spectra for a continuous wind. For simplicity, we assume that the central engine wind has a constant luminosity and baryon loading rate, and hence, a constant Lorentz factor:
\begin{eqnarray}
 L_w (\hat t) & = & L_0, \nonumber \\
 \dot M (\hat t) & = & \dot M_0, \nonumber \\
 \eta(\hat t) & = & \Gamma(\hat t) = \Gamma_0,
\label{eq:const}
\end{eqnarray}
where $\hat t$ denotes the central-engine time since the injection of the very first layer of the wind.

In order to calculate the emission from the entire wind, we dissect the wind into many thin layers, with each layer denoted by its injection time $\hat t$. Repeating the excise discussed in \S\ref{sec:bb}, we can write the contribution of specific flux at the observer time $t$ for a layer ejected during the time interval from $\hat t$ to $\hat t + d \hat t$ (for $\hat t < t$)
\begin{eqnarray}
\hat F_{\rm \nu}(\nu,t,\hat t) &  = & {\dot N_0 (\hat t) \over 4 \pi d_{\rm L}^2} \int \int P(r,\mu) P(\nu,T)h\nu \nonumber \\
& \times & \delta \left(t - \hat t - {r u \over \beta c} \right)d \mu dr.
\label{eq:Fnuhat}
\end{eqnarray}
The $\delta$-function here takes into account the retardation effect for different layers ejected at different injection time $\hat t$. The parameter $\dot N_0(\hat t)$ is the instantaneous injection rate of photons at the central engine time $\hat t$, and a rough normalization
\begin{equation}
\dot N_0(\hat t)={L_w (\hat t) \over k T(\hat t)}
\label{eq:Ndot}
\end{equation}
is adopted. The calculation of $T(\hat t)$ follows Eqs. (\ref{eq:comoving_T}) and (\ref{eq:T}), with $L_0$ replaced by $L_w(\hat t)$. Notice that the parameter $\hat F_\nu$ has the dimension of specific flux over time.

The total observed instantaneous specific flux at $t$ can be obtained by integrating $\hat F_\nu$ over all the layers, i.e.

\begin{equation}
F_{\nu}(\nu,t)=\int^t _0 \hat F_{\nu}(\nu,t,\hat t) d \hat t.
\label{eq:Fnu2}
\end{equation}

\begin{figure}[ht]
\includegraphics[angle=0,scale=0.45]{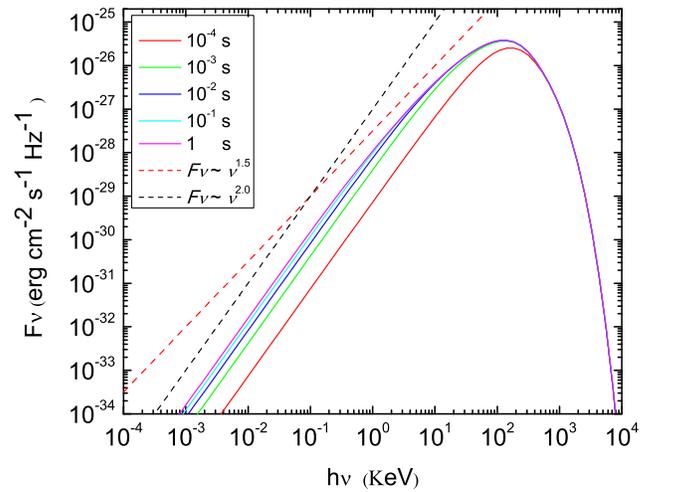}
\caption{The instantaneous photosphere spectra of a continuous wind. The parameters are the same as Fig.\ref{fig:bb} except $E_0$ is not used and $L_w(\hat t) \equiv L_0=10^{52}~{\rm erg~s^{-1}}$ is adopted. The two dashed lines are the reference lines for $F_\nu \propto \nu^2$ (black) and $F_\nu \propto \nu^{1.5}$ (red), respectively.}
\label{fig:wind}
\end{figure}

\begin{figure}[ht]
\includegraphics[angle=0,scale=0.45]{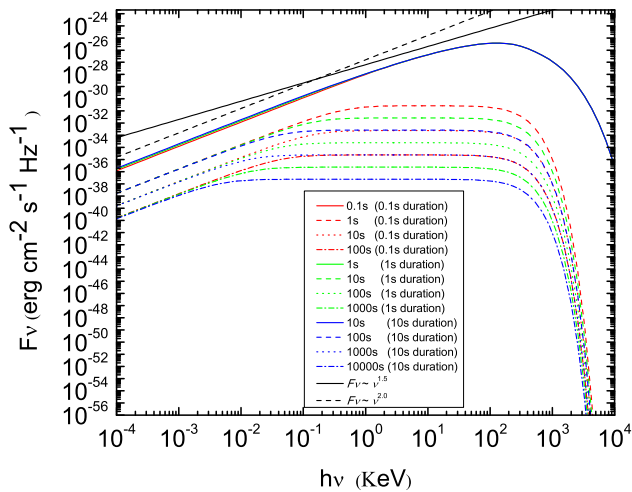}
\caption{The instantaneous photosphere spectra of a continuous wind, which shuts down at a particular time. Parameters are same as Fig.\ref{fig:wind}. Different color groups represent the spectra for different shut-down times: 0.1 s (red), 1 s (green) and 10 s (blue). For each group, four instantaneous spectra with different observation times are plotted: solid (end of continuous wind), dashed (one order of magnitude after), dotted (two orders) and dash-dotted (three orders). The two black lines are the reference lines for spectral indices being 2 and 1.5.}
\label{fig:wind_shut_down}
\end{figure}

\begin{figure}[ht]
\includegraphics[angle=0,scale=0.45]{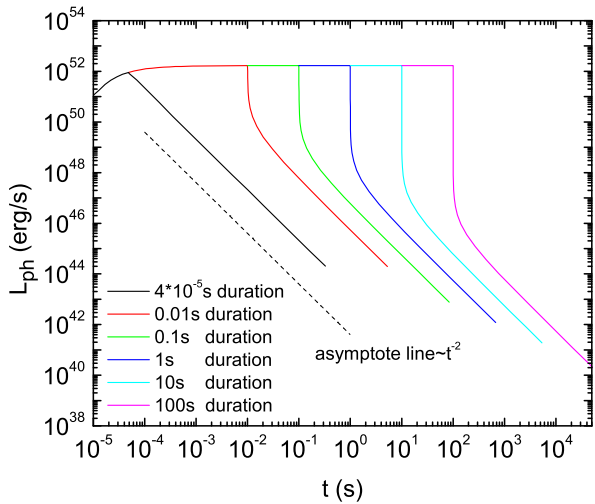}
\caption{Photosphere luminosity light curves for continuous winds with abrupt shut-down of the central engine. Different colors represent different shut-down times. For the cases with duration longer than the characteristic duration ($t_{_{\rm N}}\sim 4\times 10^{-5}$ s), the light curves initially fall rapidly before entering the $t^{-2}$ phase. The longer the central engine time, the more significant the rapid drop is.}
\label{fig:wind_Lph}
\end{figure}

We study two cases in the following. In the first case, we assume that the central engine continuously injects an outflow with a constant luminosity during all the observation times (Fig.\ref{fig:wind}). In the second case, we introduce shut-down of the central engine after a certain duration time (Fig.\ref{fig:wind_shut_down} and Fig.\ref{fig:wind_Lph}). All the parameters are the same as the ones adopted in \S\ref{sec:bb}, except $E_0$ is no longer used, and $L_w(\hat t) \equiv L_0 = 10^{52}~{\rm erg~s^{-1}}$ has been adopted.

Figure \ref{fig:wind} presents the observed instantaneous photosphere spectra of a continuous wind with a constant luminosity and Lorentz factor. One can see several interesting features. The instantaneous spectrum is initially ($t=10^{-4}$ s) blackbody-like with $F_\nu \propto \nu^2$ below the peak (Rayleigh-Jeans regime). Soon after, the spectrum below the peak starts to flatten, and a new segment with $F_\nu \propto \nu^{1.5}$ starts to merge below the peak. The reason for this softening can be understood from the results presented in Fig.\ref{fig:bb}, which delineates the time evolution of instantaneous spectra of each layer. An old layer is high-latitude dominated, so that a more extended plateau with $\hat F_\nu \propto \nu^0$ spectral segment shows up. A relatively newer layer has a shorter plateau, and the newest layer has no plateau. The superposition of emission from all these layers give rise to relatively softer spectral segment. The spectral index of this new segment ($F_\nu \propto \nu^{1.5}$) is consistent with the result of \cite{beloborodov10}, who obtained a similar spectral index using a different method. Notice that in Fig.\ref{fig:wind} the absolute flux increases with time. This is because at early epochs, the outmost layer only reaches a certain $r$ above which no photons are released. Given a simplistic probability function defined in Eq.\ref{Pr} or Eq.\ref{eq:Pru_Bel}, a good fraction of photons do not contribute to the observed flux. In the following (\S\ref{sec:variable_boundary}), we will give a more accurate treatment on this effect.

Figure \ref{fig:wind_shut_down} presents the instantaneous photosphere spectral evolution with the assumption that central engine shuts down after a certain duration. We calculate three different central engine durations: 0.1s (red group curves), 1s (green group curves) and 10s (blue group curves). For each case, we calculate four instantaneous spectra with different observational times: solid line (end of constant luminosity injection), dashed line (one order of magntitude in time after the injection phase), dotted line (two orders of magnitude after) and dash-dotted line (three orders of magnitude after). The results show that the shape of the spectrum become high-latitude dominated at later times, but early on there is a rapid falling phase. In order to fully reveal this feature, we calculate the photosphere luminosity evolution with time (light curve) as shown in Fig.\ref{fig:wind_Lph}. Our results show that the luminosity evolution depends on the duration of the central engine. Even though at late times the decaying slope of $L_{\rm ph}$ is $t^{-2}$, shortly after wind terminates, $L_{\rm ph}$ decays rapidly like free-fall\footnote{Notice that the calculated photosphere luminosity slightly deviates from the input wind luminosity $L_0=10^{52}~{\rm erg~s^{-1}}$. This is caused by the inaccurate estimate of the normalization parameter $\dot N$ (Eq.(\ref{eq:Ndot}).}. The longer of the central engine duration, the more significant the initial rapid drop it is. Only when the duration becomes as short as a characteristic duration $t_{_{\rm N}}$\citep{peer11} $\sim r_{ph} / (2 \Gamma^2 c) \sim 4\times 10^{-5}$s, does the rapid falling phase disappear (black curve in Fig.\ref{fig:wind_Lph}. This feature is caused by the ``initial time effect" for $\log - \log$ plots (see also Figure 3 of \cite{zhang06} in the case of afterglow emission). Previously \cite{ryde09} analyzed the data from 56 long GRBs. They found that the light curves decay rate is universally around $t^{-2}$. They considered this as consistent with the prediction of the high latitude emission of the photosphere model. Our results in Fig.\ref{fig:wind_Lph} suggest that this interpretation is unlikely, since there is no steep decay phase (4 orders of magnitude drop in flux for a 1 s wind) observed.

\subsection{Continuous wind with a constant wind luminosity, variable finite outer boundary}\label{sec:variable_boundary}

So far we have assumed that the outer boundary of the outflow is at infinity. For a constant luminosity wind, the optical depth at a certain position in the outflow is time independent. However, in real situations invoking a short, variable wind from a GRB, the outer boundary is time variable. As a result, the optical depth at a certain position in the outflow is time dependent and changes rapidly with time at early epochs due to the relativistic motion of the outflow, especially during the early phase of wind injection. Since the optical depth is one of the key factors to decide the photosphere probability function, the probability function also becomes time dependent. This also affects the observed photosphere temperature, and the observed $E_p$ evolution in the photosphere model.

Technically, since our model is limited to the $r_{\rm ph}>r_{\rm s}$ case, in our calculation we keep track the evolution of $r_{\rm ph}$ and compare it with $r_{\rm s}$, to make sure the presented results are relevant ones when $r_{\rm ph}>r_{\rm s}$ is satisfied. The assumption of constant luminosity and Lorentz factor (Eq.\ref{eq:const}) is still adopted in the calculations.

\subsubsection{Optical depth calculation}\label{subsubsec:optical_depth}

For a wind with boundary at infinity, the optical depth can be written as \citep{abramowicz91, beloborodov11}:
\begin{eqnarray}
\tau&=&\int_{r_1}^{\infty} d\tau \nonumber\\
&=&\int_{r_1}^{\infty} {\cal D}^{-1}\sigma_{\rm T} n' ds\nonumber\\
&=&\int_{r_1}^{\infty} {\cal D}^{-1}\sigma_{\rm T} n' dr/\cos\theta,
\label{eq:int_tau_B1}
\end{eqnarray}
where $r_1$ is the photon emission radius, $ds$ and $dr$ are along the ray direction and radial direction, respectively, ${\cal D}=[\Gamma(1-\beta \mu)]^{-1}$ is the doppler factor and
\begin{equation}
n^\prime=\frac{\dot{M}}{4\pi m_{\rm p} \beta c \Gamma r^2}
\label{eq:n'}
\end{equation}
is the number density in the comoving frame. Since the number density in rest frame is
\begin{equation}
n=\Gamma n^\prime,
\label{eq:n}
\end{equation}
the above equation can be written as
\begin{eqnarray}
\tau&=&\int_{r_1}^{\infty} \Gamma(1-\beta \mu)\sigma_{\rm T} n' dr/\cos\theta \nonumber\\
&=&\int_{r_1}^{\infty} (1-\beta \mu)\sigma_{\rm T} n dr/\cos\theta.
\label{eq:int_tau_B2}
\end{eqnarray}

\begin{figure}
\includegraphics[angle=0,scale=0.50]{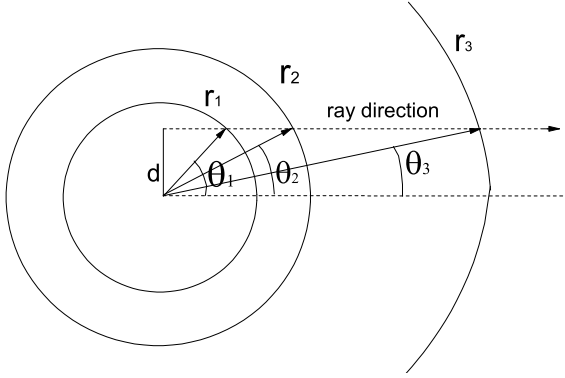}
\caption{Geometric relations of the catch-up process.}
\label{fig catch_up_1}
\end{figure}

For an finite outer boundary we are considering, the upper limit of integration has to be modified to a finite value. Assuming that a photon is emitted at a position ($r_1,\theta_1$), ahead of which there is a shell of materials extending to an outer boundary at $r_2$ (Fig.\ref{fig catch_up_1}). The light ray intersects with the out boundary of the shell at ($r_2,\theta_2$). Since the shell is also expanding near speed of light, the location as the photon catches up with the front of the shell would be at ($r_3,\theta_3$). The optical depth should then be calculated by
\begin{eqnarray}
\tau&=&\int_{r_{\rm 1}}^{r_{\rm 3}} d\tau \nonumber\\
%&=&\int_{r_{\rm 1}}^{r_{\rm 3}} \sigma_{\rm T} n dr'/\cos\theta\nonumber\\
&=&\int_{r_{\rm 1}}^{r_{\rm 3}} (1-\beta \mu)\sigma_{\rm T} n dr/\cos\theta,
\label{eq:int_tau_D1}
\end{eqnarray}
where $r_3$ can be solved through the equations
\begin{equation}
r_3 \cos\theta_3-r_1 \cos\theta_1=c \Delta t_{\rm c},
\label{eq:motion_photon}
\end{equation}
and
\begin{equation}
r_3 -r_2=\beta c \Delta t_{\rm c},
\label{eq:motion_outflow}
\end{equation}
where $\Delta t_{\rm c}$ is the time for the photons emitted from ($r_1,\theta_1$) to catch up with the outer boundary at ($r_3,\theta_3$).
Meanwhile, a simple geometrical formula gives
\begin{equation}
d=r_1 \sin\theta_1=r_2 \sin\theta_2=r_3 \sin\theta_3,
\end{equation}
where $d$ is the distance between the line of ray and the axis of the explosion along line-of-sight (Fig.\ref{fig catch_up_1}).

These equations can be solved in two different ways. First, we can calculate the catching-up outer boundary position $r_3$ based on the initial value of $r_1$, $\theta_1$ and $r_2$. The solution is:
\begin{equation}
r_3=\Gamma^2(-{\cal A}+\sqrt{{\cal A}^2-({\cal A}^2+\beta^2 d^2)/\Gamma^2},
\label{eq:r_3}
\end{equation}
where ${\cal A}=\beta r_1 \cos\theta_1 - r_2$. Here only one physical solution of $r_3$ (two mathematic solutions) is kept.

Second, one can solve for $r_2$ using $r_3$, i.e.
\begin{eqnarray}
r_2&=&r_3(1-\beta \cos\theta_3)+\beta r_1 \cos\theta_1,\nonumber\\
&=&r_3(1-\beta \sqrt{1-{d^2 \over {r_3}^2}})+\beta r_1 \cos\theta_1,\nonumber\\
&=&r_3-\beta \sqrt{{r_3}^2-d^2}+\beta r_1 \cos\theta_1.
\label{eq:r_2}
\end{eqnarray}
By employing $r_2 = r_1 + \beta c \delta {\hat t}$, one can find out $\delta {\hat t}$, which is the emission time difference between layers at position $r_2$ and $r_1$. This second approach is applied during integration when a variable wind luminosity is introduced (see details in \S\ref{sec:variable_luminosity}).

\subsubsection{Modified probability function}\label{sec:P}

With a finite outer boundary, one needs to modify the probability function of last scattering from the simple form with the infinite boundary (Eqs.(\ref{Pr}-\ref{eq:Pru_Bel})). In this sub-section we develop a general method to calculate the probability function.

We first recall a simple radiation transfer model: $I=I_{\rm 0}e^{\rm -\tau}$, where $I_{\rm 0}$ is the initial radiation intensity and $I$ is the observed intensity after absorbtion (scattering in the current case) with optical depth $\tau$. So $I/I_{\rm 0} = e^{\rm -\tau}$ (which is $\sim \tau$ when $\tau \ll 1$) is the fraction of the remaining radiation flux, which would stand for the probability of not being scattered. The factor $1-I/I_{\rm 0}=1-e^{-\tau}$, on the other hand, stands for the probability of being scattered.

The probability function for last scattering can be calculated in three steps. First, the probability for a photon being scattered from radius $r$ to $r+dr$ should be
\begin{equation}
P_{\rm r} dr \propto d\tau= \sigma_{\rm T} n(r) dr.
\end{equation}
Second, the probability for the photon to be scattered to the observer's direction can be expressed as
\begin{equation}
P_{\Omega} d\Omega \propto -{{\cal D}^2 d\mu d\phi \over 4\pi}={{\cal D}^2 d\Omega \over 4\pi}.
\end{equation}
Here we have noticed that in the comoving frame of the flow, the probability to have the photon scattered to any direction is random, so that
\begin{equation}
P'_{\Omega'} d\Omega' \propto {d\Omega' \over 4\pi}={\sin\theta' d\theta' d\phi' \over 4\pi}=-{d\mu'd\phi' \over 4\pi},
\end{equation}
$P'_{\Omega'} d\Omega'=P_{\Omega} d\Omega$, $d\mu'={\cal D}^2 d\mu$, and $d\phi'=d\phi$.
Finally, the probability for this scattered photon not being scattered again is $e^{-\tau}$.
Putting everything together, one can write
\begin{equation}
P(r,\Omega)={\sigma_{\rm T} n {{\cal D}^2 } e^{-\rm \tau(r,\mu,r_{\rm out})
} \over 4 \pi A}
\label{eq:PrOmega}
\end{equation}
where the normalization factor is
\begin{eqnarray}
A&=&\int\int P(r,\Omega) dr d\Omega \nonumber\\
&=&\int_{\rm r_{min}}^{\rm r_{max}}\int_{0}^{1}\int_0^{\rm 2\pi} \sigma_{\rm T} n {{\cal D}^2 \over 4\pi} e^{-\rm \tau(r,\mu,r_{\rm out})} dr d\mu d\phi \nonumber\\
&=&\int_{\rm r_{min}}^{\rm r_{max}}\int_{0}^{1} \sigma_{\rm T} n {{\cal D}^2 \over 2} e^{-\rm \tau(r,\mu,r_{\rm out})} dr d\mu~.
\label{eq:A}
\end{eqnarray}
Here the function $\tau(r,\mu,r_{\rm out})$ is a function depending on $r$ and $\mu$, and the outer boundary $r_{\rm out}$ of the outflow at the time when the photon crosses the ejecta. It is $r$-dependent, and $\hat t$-dependent. The relation between $r$ and $r_{\rm out}$ is the same as $r_1$ and $r_3$ discussed above (\S\ref{subsubsec:optical_depth}). Since this function is rapidly evolving with time, the probability function (Eq.(\ref{eq:PrOmega})) is also rapidly evolving.

In the normalization function (Eq.(\ref{eq:A})), the integration limits $r_{\rm min}$ and $r_{\rm max}$ should be 0 and $+\infty$, respectively. In reality, we take $r_{\rm min} \sim r_0 = 10^7$ cm, and $r_{\rm max}$ a large enough number (e.g. $5\times 10^{17}$ cm). Since the probabilities at very small and very large distances are both very small. The actual numerical values of the two limits essentially do not affect the calculation results.

\begin{figure}
\includegraphics[angle=0,scale=0.45]{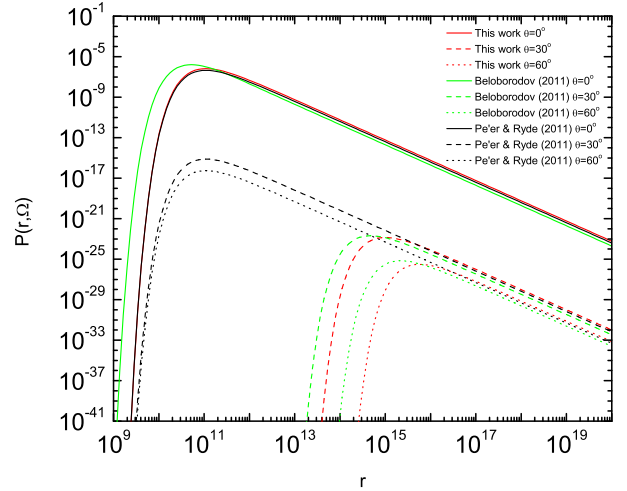}
\caption{A comparison of the probability function $P(r,\Omega)$ between this work and those of \cite{peer11} and \cite{beloborodov11}.}
\label{fig:compare-probability}
\end{figure}

In Fig.\ref{fig:compare-probability}, we compare our probability function in the infinite outer boundary limit with those of \cite{peer11} and \cite{beloborodov11}. Our results agree with \cite{peer11} in the small angle limit ($\theta=0$), and are more consistent with that of \cite{beloborodov11} in the large angle limit ($\theta=30^{\rm o},60^{\rm o}$).

\subsubsection{Putting pieces together}

The specific flux at time $t$ for a layer ejected at the time $\hat t$ can be expressed as
\begin{eqnarray}
\hat F_{\nu}(\nu,t,\hat t)&=& {\dot N_0(\hat t) \over 4 \pi d_{\rm L}^2} \int \int \hat P(r,\Omega) P(\nu,T)h\nu \nonumber\\
&&\times \delta \left(t - \hat t - \left({r u \over \beta c}-t_{0}\right) \right)d\Omega dr.
\label{eq:f_nu_2_3_1}
\end{eqnarray}
Compared with the calculation in \S\ref{sec:wind} (Eq.(\ref{eq:Fnuhat})), this expression has two improvements. First, we have introduced $t_{0}={r_0 \over \beta c (1+\beta) \Gamma^2}$ to reflect that the wind is ejected from a central engine with radius $r_0=10^7$ cm (instead of $r_0 = 0$). Second, since $P(r,\Omega)$ is no longer a universal function, we express it as $\hat P(r,\Omega)$ to specify that it is related to the layer ejected at $\hat t$. The expanded full expression of $\hat F_{\nu}(\nu,t,\hat t)$ (Eq.(\ref{eq:f_nu_2_3_1})) is presented in the Appendix.

Again the instantaneous specific flux can be calculated by integrating the contributions from all layers
\begin{equation}
F_{\rm \nu,}(\nu,t)=\int^t _0 \hat F_{\rm \nu}(\nu,t,\hat t) d \hat t.
\label{eq:f_nu_2_3_2}
\end{equation}

Finally, one can integrate over a time interval ${\rm [t_1,t_2]}$ to get a time integral spectrum, which is what is observed:
\begin{equation}
F_{\rm \nu}(\nu,t_1\rightarrow t_2) = \int_{\rm t_1}^{\rm t_2} F_{\rm \nu}(\nu,t)~ dt.
\label{eq:f_nu_2_3_3}
\end{equation}

\subsubsection{Results}

Here, we present the calculation results for a constant wind luminosity with a variable outer boundary. The input parameters are: constant wind luminosity $L_{\rm w}=10^{52}~ {\rm erg~s^{-1}}$, constant dimensionless entropy $\eta(\hat t)=\Gamma(\hat t)=\Gamma_0=300$, luminosity distance of the central engine $d_{\rm L}=2\times 10^{28}$ cm $(z \sim 1)$, and central engine radius $r_0=10^7$ cm.

\begin{figure}[ht]
\includegraphics[angle=0,scale=0.45]{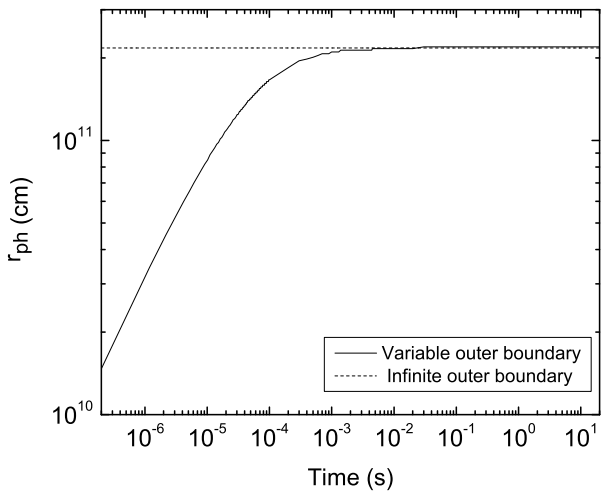}
\caption{The evolution of $r_{ph}$ as a function of observer time for a constant luminosity central engine wind. The dotted line is the asymptotic solution for an outer boundary at infinity.}
\label{fig_r_ph_L_const}
\end{figure}

\begin{figure}[ht]
\includegraphics[angle=0,scale=0.45]{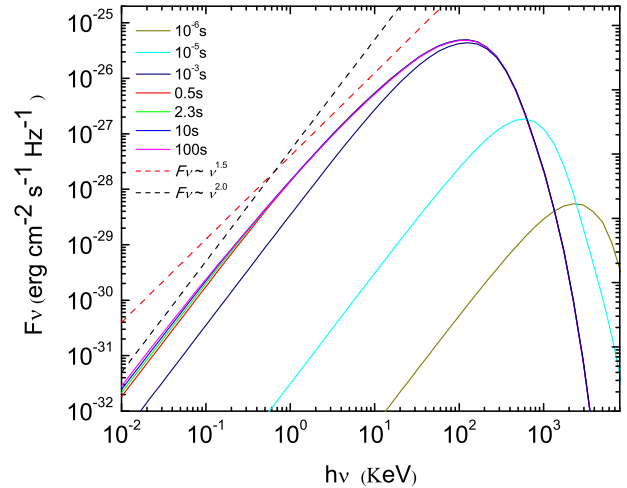}
\caption{The instantaneous spectra for a constant luminosity wind with a variable outer boundary. Different colors stand for different epochs: $10^{-6}$ s (dark yellow), $10^{-5}$ s (cyan), $10^{-3}$ s (navy), 0.5 s (red), 2.3 s (green), 10 s (blue), and 100 s (magenta). }
\label{fig:L_const_spectrum}
\end{figure}

Since our spectral calculation is valid for $r_{ph}>r_{\rm s}$, we need to closely track the location of $r_{ph}$. Figure \ref{fig_r_ph_L_const} shows the numerical results of $r_{\rm ph}$ evolution ($L_w=10^{52}~{\rm erg~s^{-1}}$ and $\Gamma_0=300$). It shows that before $10^{-3} s$, $r_{\rm ph}$ increases rapidly, while after $10^{-3} s$, $r_{ph}$ is nearly constant around $2\times10^{11}$ cm. Since $r_{\rm s}=\eta r_0=3\times10^9$ cm, the condition $r_{\rm ph}>r_{\rm s}$ is easily satisfied from very early time. We choose seven observer times, $10^{-6}, 10^{-5}, 10^{-3}, 0.5, 2.3, 10, 100$ seconds, to calculate the instantaneous spectra. The results are shown in Figure \ref{fig:L_const_spectrum}. We find that early on ($10^{-6}-10^{-3}$ s, the spectra evolve rapidly. In particular, the temperature displays a strong hard-to-soft evolution. This is because initially the photosphere radius is closer in due to the less opacity early on (Fig.\ref{fig_r_ph_L_const}). However, such a phase is too short to have an observational consequence. After $10^{-3}$ s, the photosphere radius approaches the sympototic value, so is the photosphere temperature. As a result, the last four spectra are nearly identical with minor differences in the low energy regime. In other words, there is essentially no temporal evolution with time. For a constant initial wind luminosity, the density and probability function for different layers are essentially the same. As a result, given a same $r_{ph}$ the spectral behavior is rather similar. From the results, we find that the spectral index from $3$ keV to the peak is modified from 2 to $\sim 1.5$. This is again due to the superposition from the older layers' high latitude contribution, as already analyzed in \S\ref{sec:wind}. Compared with the results presented in \S\ref{sec:wind} (which has an outer boundary at infinity), the more sophisticated method here can trace the evolution of the photosphere radius and probability function for a dynamic, finite outer boundary. For the constant luminosity case discussed in this sub-section, the finite boundary treatment makes a noticeable difference only at very early times ($t<10^{-3} s$). The difference is more obvious for a variable luminosity wind, as we discuss next.

\subsection{Variable wind luminosity}\label{sec:variable_luminosity}

In a real GRB, wind luminosity varies rapidly with time. Unlike the constant luminosity/$\Gamma$ model which preserves a $n \propto r^{-2}$ density profile of the wind, the density profile changes rapidly with time depending on the time history of luminosity, $L_w(\hat t),$ and baryon loading, $\dot M(\hat t)$. The integration for optical depth (Eq.(\ref{eq:int_tau_D1})) becomes more complicated. In this sub-section, we develop a method to handle this problem. For simplicity, we assume a power law form of the time history, and assume a constant $\Gamma = \eta = L_w(\hat t)/\dot M(\hat t) c^2$ throughout.

\subsubsection{Luminosity and baryon-loading history}\label{subsubsec:L_profile}

We approximate a GRB pulse as broken power law in luminosity, with rising and decaying indices as $a_r$ and $a_d$, respectively, with a peak $L_{w,p}$ at $\hat t_p$. The luminosity history in the rising phase ($\hat t < \hat t_p$) can be written as
\begin{equation}
\log L_{\rm w}(\hat t)= a_r \log \hat t + b_r,
\label{eq:Lwr}
\end{equation}
while that in the decaying phase ($\hat t > \hat t_p$) can be written as
\begin{equation}
\log L_{\rm w}(\hat t)= a_d \log \hat t + b_d,
\label{eq:Lwd}
\end{equation}
where $b_r = \log L_{w,p} - a_r \log \hat t_p$ and $b_d = \log L_{w,p} - a_d \log \hat t_p$ are normalization parameters of the two power law segments.

\subsubsection{Complications in the catch-up process}\label{sec:catchup}

For the constant $L_w$ and $\dot M$ wind case as discussed in \S\ref{sec:variable_boundary}, since the $n$ profile does not evolve with time, the integration of (\ref{eq:int_tau_D1}) from $r_1$ (photon emission radius) to $r_3$ (the radius where the photon escape the wind) is straightforward (see Fig.\ref{fig catch_up_1} and \S\ref{subsubsec:optical_depth} for detailed discussion). One can apply Eq.(\ref{eq:r_3}) to directly solve for $r_3$ given any $r_2$, and the final optical depth is defined by the maximum catching up radius $r_{\rm 3,M}$ (which is also $r_{\rm out}$ discussed in \S\ref{sec:P}) corresponding to the maximum $r_{\rm 2,M}$ of the wind at the time when the photon is emitted at $r_1$.

In the variable wind case discussed here, since the $n$ profile is variable with time, one needs to precisely determine $n$ at any $r_3$ value in the range of $(r_1,r_{\rm 3,M})$. To do so, one needs to find out the corresponding $r_2$ of each $r_3$ using Eq.(\ref{eq:r_2}). Then one can connect $r_2$ with a certain central engine time $\hat t$, and hence, its baryon loading rate $\dot M(\hat t)$, which defines the density profile, and hence, the relevant $n$ at that $r_3$.

\begin{figure}
\includegraphics[angle=0,scale=0.45]{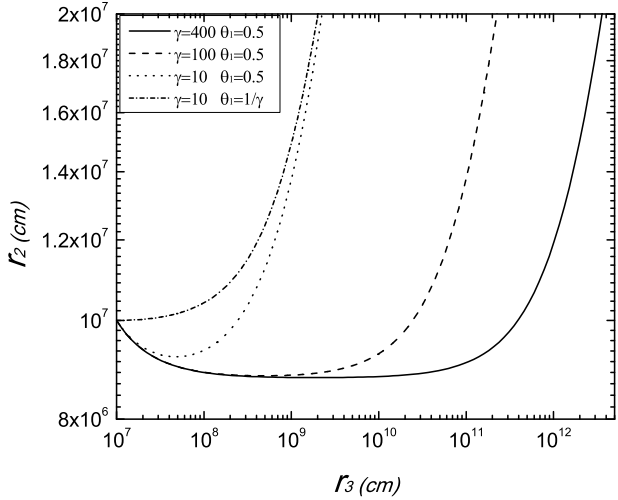}
\caption{The relationship between $r_2$ and $r_3$ based on Equation (\ref{eq:r_2}). The solid, dashed, and dotted lines have the same initial angle $\theta_1=0.5$ but different Lorentz factors $\Gamma_0=$400, 100 and 10. The dash-dotted line corresponds to the case of $\theta_1=1/\Gamma_0$ with $\Gamma_0=10$ .}
\label{fig_r2_r3_evo}
\end{figure}

Solving for Eq.(\ref{eq:r_2}), we find an interesting fact (Fig.\ref{fig_r2_r3_evo}): at low latitudes with respect to the line-of-sight (small $\theta_1$), $r_2$ is always larger than $r_1$. This means that the emitted photons always catch up layers ejected earlier. However, at large latitudes (large $\theta_1$), $r_2$ can be smaller than $r_1$. This means that photons emitted at a certain epoch would initially interact with the layers ejected later, so that it would see an even higher optical depth during propagation. This does not mean that the late ejected materials move with a superluminal velocity.  A photon is caught up with by the electrons ejected later due to a geometric effect: electrons move in a hypotenuse ``short-cut'', even though their bulk motion velocity is sub-luminal (Fig.\ref{fig catch_up_1}).

For a large-angle geometry when the reverse of $r_2$ and $r_1$ happens, the relationship between $r_2$ and $r_3$ shows an interesting feature. Performing derivative to equation (\ref{eq:r_2}), one gets
\begin{eqnarray}
{dr_2 \over dr_3}=1-{\beta r_3 \over \sqrt{{r_3}^2-d^2}}=1-{\beta \over \cos\theta_3}.
\label{eq:dr2_dr3}
\end{eqnarray}
Setting $dr_2/dr_3=0$, one gets a critical catch-up point $\cos\theta_3=\beta$, or $\sin\theta_3=1/\Gamma_0$. In the early stages, $\theta_3$ is relatively large, so that $\sin\theta_3>1/\Gamma_0$ and $dr_2/dr_3<0$ are satisfied. As a result, initially $r_2$ is smaller than $r_1$ (e.g. $\sim 10^7$ cm), and $r_2$ decreases when $r_3$ increases. After passing the critical point (Eq.\ref{eq:dr2_dr3}), $dr_2/dr_3$ becomes positive. The photon starts to overtake the outflow layers and eventually escape the wind. If the initial angle $\theta_1$ is small enough from the beginning, i.e. $\sin\theta_1 < 1/\Gamma_0$ is satisfied from the beginning, $r_2$ would increase with $r_3$ all the time. In Figure (\ref{fig_r2_r3_evo}), we show four different cases of $r_2 - r_3$ evolution. The solid, dashed, and dotted lines have the same initial angle $\theta_1=0.5$ but with different Lorentz factors. The dash-dotted line corresponds to the critical case $\theta_1=1/\Gamma_0$.

\subsubsection{Optical depth calculation}

\begin{figure}
\includegraphics[angle=0,scale=0.45]{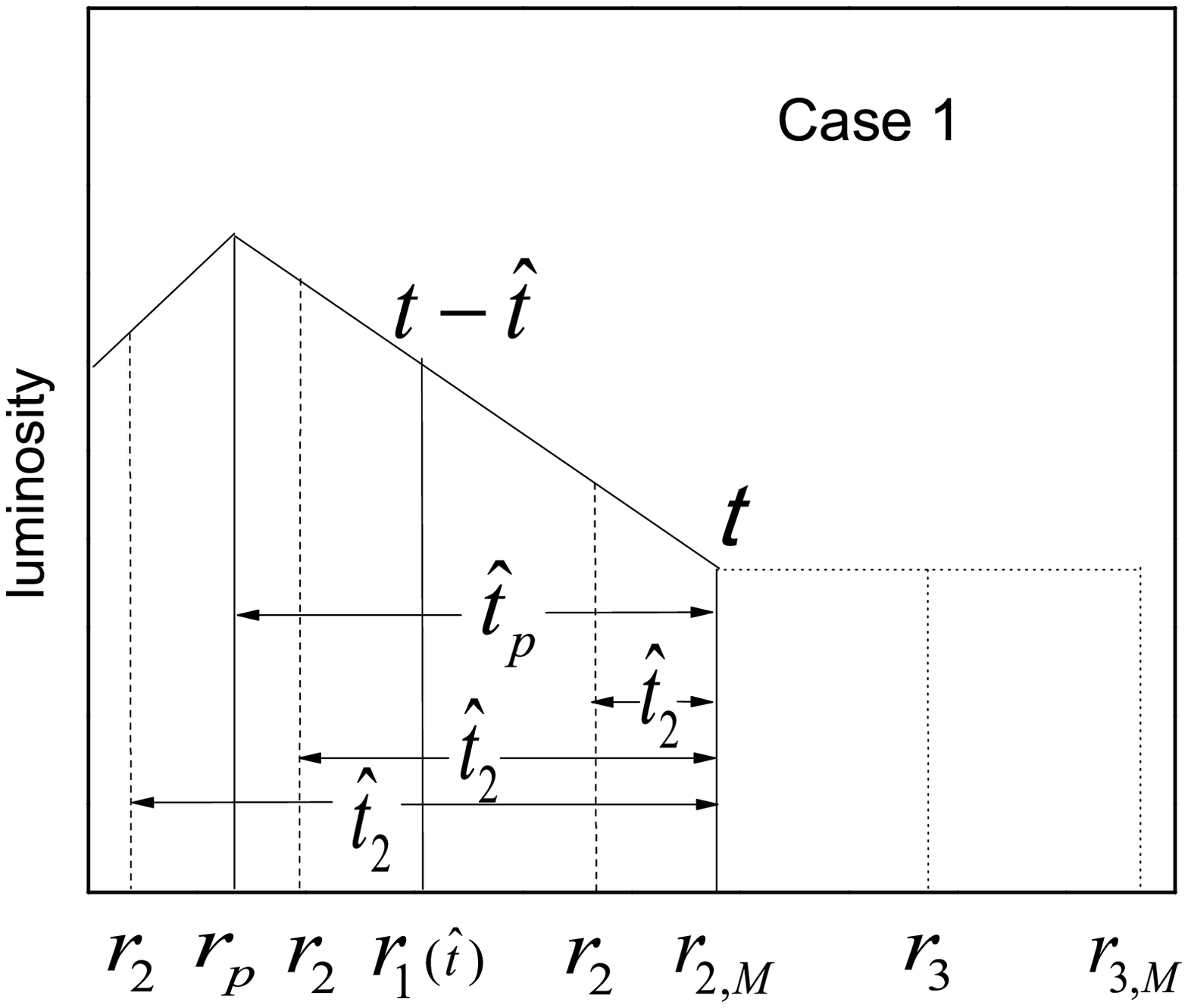}
\includegraphics[angle=0,scale=0.45]{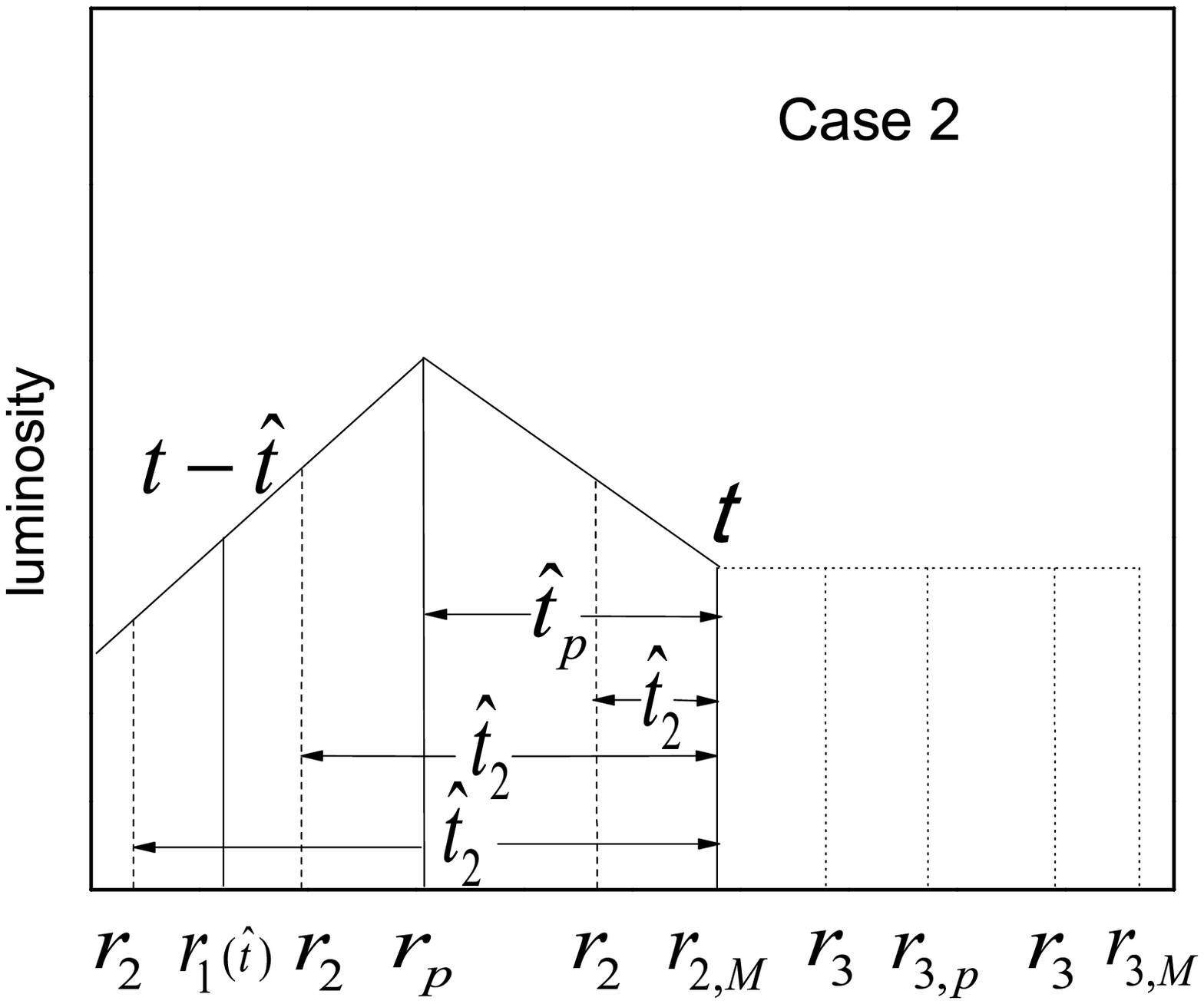}
\caption{Instantaneous spacial structure of the wind luminosity as well as typical radii invoked to calculate optical depth.}
\label{fig 3 1}
\end{figure}

With the above preparation, one can calculate optical depth of a photon in a variable-luminosity wind. Figure \ref{fig 3 1} shows the spacial distribution of wind luminosity at an instant $t$ in the observer frame. The layers ejected at an earlier epochs move in the front, so the spacial distribution is essentially a reversed temporal distribution. We have assumed a constant Lorentz factor in all layers so that the temporal profile does not evolve with time other than globally streaming forward.

We calculate the optical depth of a photon emitted from a layer ejected at the central engine time $\hat t$, which has an age of $t-\hat t$ at the observer time $t$. The radius of the layer is defined as $r_1$ (or more precisely $r_1(\hat t)$). We discuss two cases in Fig.\ref{fig 3 1}: ``Case 1'' corresponds to the early stage when the photon is emitted during the rising phase of the pulse, while ``Case 2'' corresponds to the late stage when the photon is emitted during the falling phase. In both cases, the location of the layer emitted at the peak time $\hat t$ is denoted as $r_p$.

At the time $t$ when a photon is emitted at $r_1(\hat t)$, all the layers emitted before $\hat t$ are ahead of $r_1$, and the maximum radius $r_{\rm 2,M}$ is the first layer ejected at $\hat t = 0$. The photon will over-take this layer at a much larger radius $r_{\rm 3,M}$. In order to calculate the optical depth of the photon $\tau$, one needs to integrate Eq.(\ref{eq:int_tau_D1}) from $r_1$ to $r_{\rm 3,M}$. For any $r \in (r_1, r_{\rm 3,M})$ ($r$ is effectively $r_3$), the density $n(r)$ is defined by its corresponding $r_2$, which is related to $\dot M (\hat t_2)$ at the time
\begin{equation}
\hat t_2 = \frac{r_{\rm 2,M} - r_2}{\beta c},
\end{equation}
when the layer associated with $r_2$ was ejected. One therefore needs to solve for $r_2$ using Eq.(\ref{eq:r_2}) for every step in $r$ (i.e. $r_3$).

In view of the complicated catch-up process discussed in \S\ref{sec:catchup}, there are three possibilities for each case (Fig.\ref{fig 3 1}). For Case 1, one has $r_p < r_1 < r_2$, $r_p < r_2 < r_1$, or $r_2 < r_p < r_1$. The wind luminosity at $\hat t_2$ is defined by Eq.(\ref{eq:Lwr}) for the first two cases, and by Eq.(\ref{eq:Lwd}) for the last case. For Case 2, one has $r_1 < r_p < r_2$, $r_1 < r_2 < r_p$, or $r_2 < r_1 < r_p$. The wind luminosity at $\hat t_2$ is defined by Eq.(\ref{eq:Lwr}) for the first case, and by Eq.(\ref{eq:Lwd}) for the last two cases. From $L_w (\hat t_2)$ one can calculate $\dot M(\hat t_2) = L_w(\hat t_2)/\eta c^2$ given the constant $\eta$ value, which can be used to calculate the density $n$ using Equations (\ref{eq:n'}) and (\ref{eq:n}). One can then complete integration to calculate the optical depth $\tau$ for any coordinate $(r,\theta)$. The photosphere radius $r_{ph}^\theta$ at any angle $\theta$ is defined by the condition
\begin{equation}
 \tau(r_{ph}^\theta,\theta) = \int_{r_{\rm ph}}^{r_{\rm 3,M}} (1-{\beta \cos \theta}) \sigma_{\rm T} n dr/\cos \theta=1.
\label{eq:int_tau_D_rph}
\end{equation}
And the traditional photosphere radius is defined as $r_{ph} = r_{ph}^{\theta=0}$.

\subsubsection{Results}

We perform calculations for several different luminosity profiles. We fix $L_{w,p} = 10^{52}~{\rm erg~s^{-1}}$, $\hat t_p = 2.4$ s, $\eta=\Gamma_0 = 300$, $r_0 = 10^7$ cm, and $d_{\rm L} = 2\times 10^{28}$ cm (effectively $z \sim 1$), and investigate six different luminosity profiles with $(a_r,a_d) = (+0.75, -1)$, $(+0.75, -2)$, $(+0.75, -5)$, $(+2, -1)$, $(+2, -2)$, $(+2, -5)$, respectively. Analytically, the on-axis ($\theta=0$) photosphere radius \citep{meszarosrees00}
\begin{equation}
r_{ph} = \frac{L_w \sigma_{\rm T}}{4\pi m_p c^3 \eta^3} \simeq 3.7\times 10^{11}~{\rm cm} L_{\rm w,52} \eta_{2.5}^{-3}
\label{eq:rph}
\end{equation}
follows the luminosity profile $L_w(\hat t)$ for a constant $\eta$, so $r_{ph}(\hat t)$ should follow the same temporal profile as $L_w(\hat t)$. We numerically reproduced this for a variable luminosity wind, with a slight deviation only at very early epochs ($\hat t < 10^{-3}$ s. The satuation radius is $r_s = \eta r_0 = 3\times 10^9~{\rm cm}~(\eta/300)$. To assure $r_{ph} > r_s$, we choose the following observed epochs to calculate the instantaneous spectra: $t=0.5$ s, 2.3 s before $\hat t_p = 2.4$ s, and $t=2.5$ s, 4 s, and 10 s after $\hat t_p$. For $a_d = -5$ case, the last epoch ($t=10$ s) already violates the $r_{ph}>r_s$ condition, so we do not include it in the calculation.

\begin{figure}[ht]
\includegraphics[angle=0,scale=0.65]{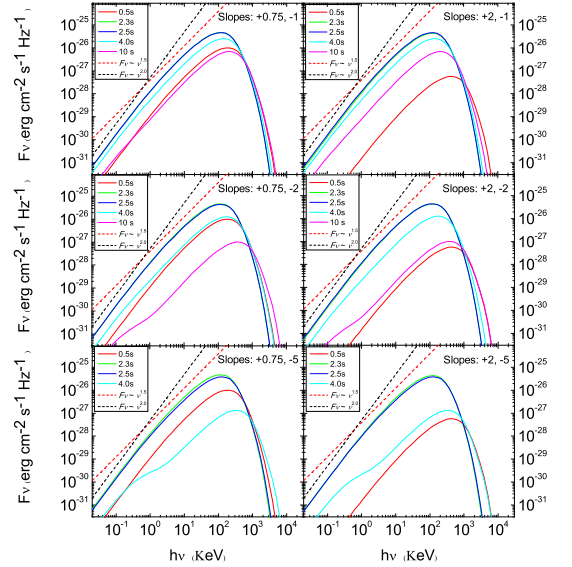}
\caption{The instantaneous photosphere spectra of winds with variable luminosity. A constant $\Gamma_0 = 300$ and $d_{\rm L}=2\times 10^{28}$ cm, a peak time $\hat t_p = 2.4$ s, and a peak luminosity $L_{w,p} = 10^{52}~{\rm erg~s^{-1}}$ are adopted for all cases. Different panels show different luminosity histories, and the temporal rising and decaying indices (slopes) are marked in each panel. For each panel, the spectra are calculated at the following times: 0.5 s (red), 2.3 s (green), 2.5 s (blue), 4 s(cyan), and 10 s (magenta). Two reference lines for spectral indices 1.5 (red dashed line) and 2 (black dashed line) are also drawn.}
\label{fig_S38_Kchange_g400_f_comb}
\end{figure}

Figure \ref{fig_S38_Kchange_g400_f_comb} shows the calculated instantaneous spectra, each panel displaying results for a luminosity profile. For each panel, different colors show different epochs. We can see that during the rising phase ($t=0.5, 2.3$ s) the resulting spectra are very similar to the cases for a constant luminosity (Fig.\ref{fig:L_const_spectrum}), i.e. the spectral slope is modified from Rayleigh-Jeans (2) to 1.5, mainly due to high-latitude contribution. During the decay phase ($t=2.5, 4, 10$ s), the spectral index below $E_p$ is somewhat shallower. This is because the high-latitude emission is more dominant since it comes from the layers that have higher luminosities. The steeper the decay phase, the more significant the high-latitude effect is.

\begin{figure}[ht]
\centering
\includegraphics[angle=0,scale=1.0]{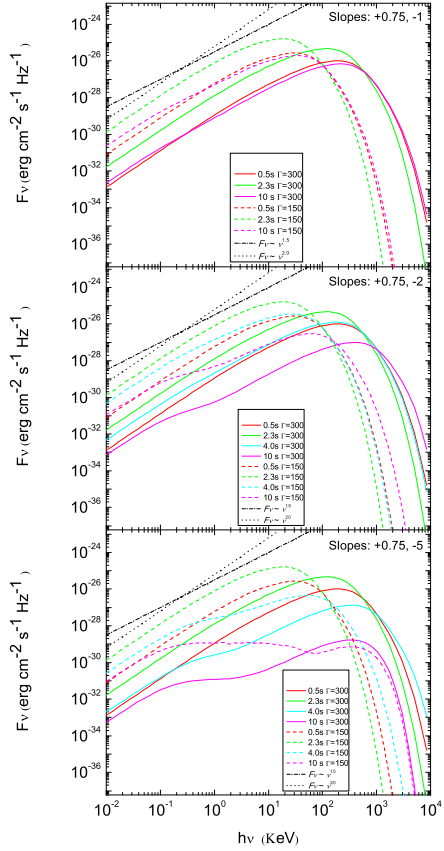}
\caption{A comparison of instantaneous photosphere spectra for different Lorentz factors: $\Gamma_0=150$ (dashed) and $300$ (solid). Other notations are the same as Fig.\ref{fig_S38_Kchange_g400_f_comb}.}
\label{fig_S38_k0.75_-1&-2&-5_g150&400}
\end{figure}

In Figure \ref{fig_S38_k0.75_-1&-2&-5_g150&400}, we compare the resulting spectra for different $\Gamma_0$. We fix $a_r = +0.75$, and vary $a_d$ for three values -1, -2, -5. For each set of luminosity profile, we compare the resulting spectra for $\Gamma_0=300$ (solid curves) and $\Gamma_0 = 150$ (dashed curves). A smaller $\Gamma_0$ corresponds to a larger photosphere radius (Eq.(\ref{eq:rph})) and a lower photosphere temperature. A larger photosphere also gives a more significant high-latitude effect, which is reflected from the somewhat shallower spectral index below $E_p$ during the decay phase, especially when $a_d$ is steep.

\begin{figure}[ht]
\centering
\includegraphics[angle=0,scale=0.45]{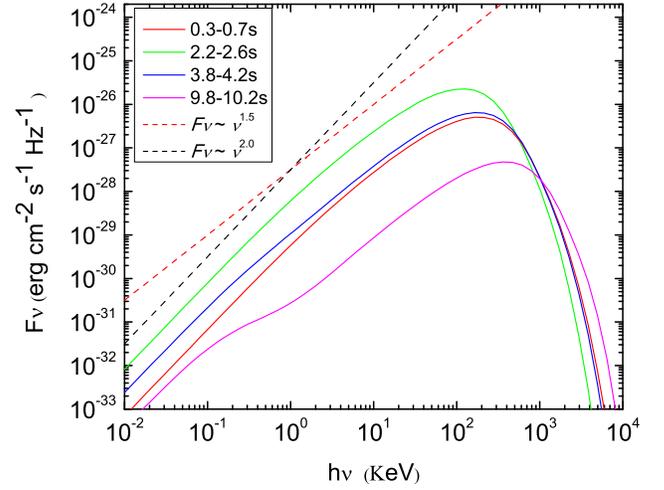}
\caption{The time integrated photosphere spectra for the case ``Slopes: +0.75, -2" with $\Gamma_0=300$. Integration in four time intervals are presented: 0.3-0.7 s (red), 2.2-2.6 s (green), 3.8-4.2 s (blue) and 9.8-10.2 s (magenta). }
\label{fig_time_integral_spectrum_bin0.5s0.1s_k0.75_-2_G400_f}
\end{figure}

Figure \ref{fig_time_integral_spectrum_bin0.5s0.1s_k0.75_-2_G400_f} shows the time integrated spectra. we choose 0.5 s as time bin for integration. For a same $\eta$ and $r_0$, the instantaneous spectra do not evolve significantly. As a result, the time-integrated spectra are not much different from the instantaneous ones.

\begin{figure}[ht]
\centering
\includegraphics[angle=0,scale=0.45]{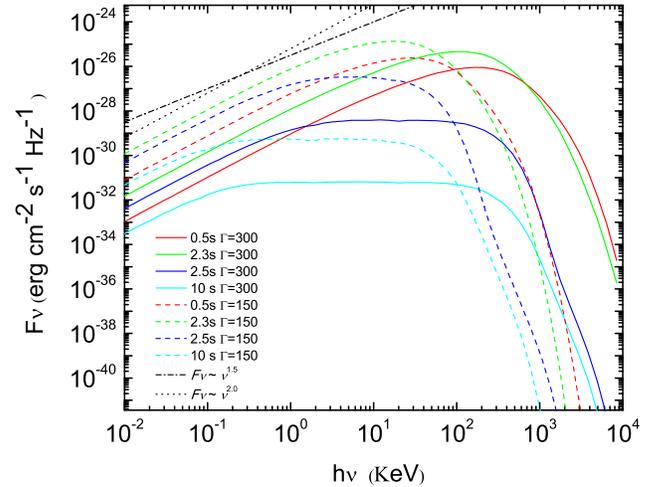}
\caption{The instantaneous photosphere spectra for a variable luminosity wind with abrupt shut-down of the central engine. The parameters are the same as Fig.\ref{fig_S38_k0.75_-1&-2&-5_g150&400}, except there is a sharp cutoff at $\hat t_p=2.4$ s. The dashed lines are for $\Gamma_0=150$, while the solid lines are for $\Gamma_0=300$. Different colors represent different observational times.}
\label{fig_S38_k0.75_-inf_g150&400_F}
\end{figure}

We also consider the case when the variable luminosity central engine wind ceases abruptly. Still keeping the same parameters, but make the wind abruptly cease at $\hat t_p = 2.4$ s, the calculation spectra are presented in Fig.\ref{fig_S38_k0.75_-inf_g150&400_F}. It is seen that at $t>\hat t_p$, a progressively more prominent plateau develops, similar to the results presented in Fig.\ref{fig:wind_shut_down}.

\subsection{$E_p$ evolution}\label{sec:Ep-patterns}

The evolution of $E_p$ is an important criterion to judge the correctness of a GRB prompt emission model. Observationally, hard-to-soft evolution and intensity-tracking patterns across a broad GRB pulse have been identified \citep{liang96,ford95,lu10,lu12}, and some tracking behavior may be due to superposition of intrinsically hard-to-soft evolution pattern in most pulses \citep{hakkila11,lu12}. It is important to check whether the quasi-thermal photosphere emission can reproduce the observed $E_p$ evolution patterns.

\begin{figure}[ht]
\centering
\includegraphics[angle=0,scale=0.45]{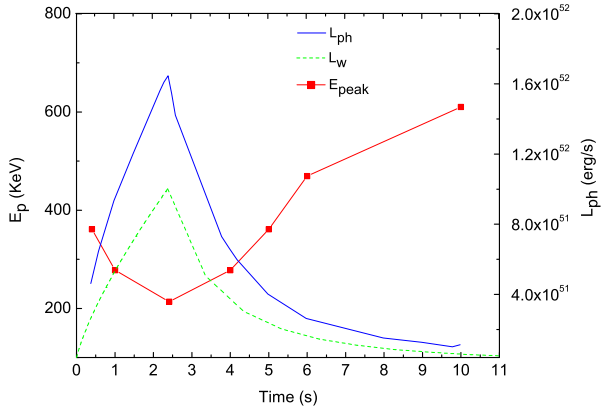}
\caption{The evolution of $E_{p}$ (Red solid line), initial wind luminosity $L_{w}$ (green dash line), and the photosphere luminosity $L_{ph}$ (blue solid line) for the case ``Slopes: +0.75, -2", $\Gamma_0=300$.}
\label{fig_Lph_Ep_Rph_Lwind_k0.75_-2}
\end{figure}

Before performing numerical calculations, it is instructive to perform some analytical estimates. For the regime $r_{ph}>r_{s}$ we are interested in, one has $\eta=\Gamma_0$, $T_{ph} \propto L_{w}^{1/4}r_0^{-1/2}{(r_{ph}/r_{s})}^{-2/3}$, $r_{s}=\eta r_0$, and $r_{ph} \sim L_{w} \eta^{-3}$, so that the observer temperature can be expressed as:
\begin{equation}
E_p \propto T_{ph} \propto L_{w}^{-5/12} r_0^{1/6} \eta^{8/3}.\
\label{eq:Ep}
\end{equation}
One can immediately see that if $\eta$ and $r_0$ are constants, $E_p$ is anti-correlated to $L_w$. This trend seems to be consistent with the ``hard-to-soft'' evolution pattern during the pulse rising phase. However, it gives an opposite trend during the pulse decaying phase, namely, $E_p$ rises as luminosity drops. Based on the numerical results of instantaneous spectra presented in Fig.\ref{fig_S38_Kchange_g400_f_comb}, we plot $E_p$ evolution with respect to wind and photosphere luminosities in Fig.\ref{fig_Lph_Ep_Rph_Lwind_k0.75_-2}. The $L_w - E_p$ anti-correlation is clearly shown. Such a pattern has never been observed in GRB pulses.

\begin{figure}[ht]
\centering
\includegraphics[angle=0,scale=0.4]{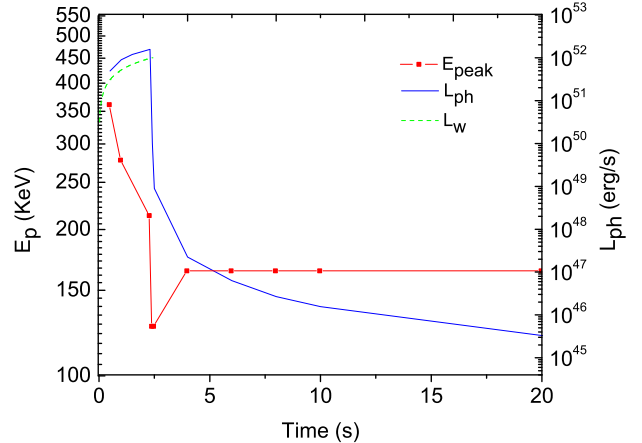}
\caption{The evolution of $E_{p}$ (Red solid line), initial wind luminosity $L_{w}$ (green dash line), and the photosphere luminosity $L_{ph}$ (blue solid line) for the case ``Slope +0.75'' with an abrupt shut-down and $\Gamma_0=300$. }
\label{fig_Lph_Ep_Rph_Lwind_k0.75_-inf}
\end{figure}

A related idea would be to attribute to the decaying phase as due to the high-latitude curvature effect. By doing so, one may expect to have $E_p$ continues to decay during the decaying phase of the pulse. Including the $E_p - L_w$ anti-correlation during the rising phase, this might reproduce the observed hard-to-soft evolution pattern. In Fig.\ref{fig_Lph_Ep_Rph_Lwind_k0.75_-inf}, based on the numerical results of the instantaneous spectra presented in Fig.\ref{fig_S38_k0.75_-inf_g150&400_F}, we plot $E_p$ evolution with respect to wind and photosphere luminosities. As can be seen from the figure, this model also cannot reproduce the data. There are two problems: First, the high-latitude curvature tail of the photosphere luminosity light curve drops rapidly (similar to Fig.\ref{fig:wind_Lph}), since the photosphere radius is small. The predicted $E_p$ evolution during the tail (even though not measurable due to the rapid decay of the flux) displays a flat feature. This is because the upper end of the flat segment of the $F_\nu$ spectrum in the high-latitude-emission-dominated phase (which defines $E_p$) essentially does not decay with time (Fig.\ref{fig:wind_shut_down}).

If one allows $\eta$ to vary with $L_w$ with a certain power law dependence, the $E_p$-evolution pattern may be modified. From Eq.\ref{eq:Ep}, one can see that if one defines $\eta=L_{w}^m$, one would have $E_p \propto L_{w}^{(-5+32m)/12}$. The $E_p - L_w$ dependence would be reversed (i.e. positive dependence) when $m > 5/32\simeq 0.156$. From afterglow data, \cite{liang10,lv12} have discovered a rough global $\Gamma_0 \propto L_w^{1/4}$ correlation in different GRBs. If such a correlation also exist within a same GRB, as theoretically motivated in GRB central engine models \citep{lei13}, then one would predict an $E_p-L$ intensity tracking behavior within this simple photosphere model. Such a pattern has been observed in a fraction of GRB pulses \citep{lu12}. For the hard-to-soft evolution case, on the other hand, in order to reproduce the data, one has to demand that the index $m$ switches from $m<5/32$ before the peak to $m>5/32$ after the peak. This requires contrived physical conditions that are not known theoretically.

\section{Conclusions and Discussion}

In this paper, we have developed a sophisticated method to calculate quasi-thermal GRB photosphere spectra numerically by introducing several improvements on previous treatments \citep{peer08,peer11}. The new ingredients introduced in this paper include: the probability distribution of the location of a dynamically evolving photosphere, superposition of emission from an equal-arrival-time ``volume'' in a continuous wind, the evolution of optical depth of a wind with finite but evolving outer boundary, as well as the effect of different wind luminosity profiles. By assuming a co-moving blackbody spectrum emerging from the photosphere and a top-hat jet profile, we address how these effects modify the observed spectra from blackbody. The following robust conclusions are drawn: 1. For an outflow with constant or increasing wind luminosity, the low-energy spectrum below $E_{p}$ can be modified to $F_\nu \sim \nu^{1.5}$, corresponding to a low-energy photon index $\alpha \sim +0.5$. Introducing temporal smearing does not change $\alpha$ significantly. 2. A softer spectrum can be obtained during the phase of decreasing wind luminosity with time, and a flat spectrum $F_\nu \sim \nu^0$ ($\alpha=-1$) can be obtained only when the spectrum is high-latitude emission dominated. However, since the photosphere radius is small, the flux drops very rapidly shortly after the wind terminates. 3. Depending on how $\eta$ is related to $L_w$, this model can give both negative or positive $E_p - L_w$ correlation. The observed ``hard-to-soft'' evolution of $E_p$ across broad pulses of seconds duration \citep{lu12} cannot be interpreted with this simple photosphere model, unless an unknown contrived physical condition to switch the index $m$ at the pulse peak is invoked. The intensity tracking patterns as observed in some broad pulses \citep{lu12} can be accounted for this model if $\eta \propto L_w^m$ with $m>5/32$.

The results presented here suggest that the observed dominant spectral component, the so-called ``Band-function'' \citep{band93} component, is not easy to interpret by this simplest photosphere model. The predicted low energy spectral index $\alpha = +0.5$) is too hard compared with the typical observed value ($\alpha = -1$), and the widely observed ``hard-to-soft'' $E_p$ evolution across broad pulses cannot be accounted for unless a contrived condition is invoked. In order to naturally interpret GRB spectra within the framework of the photosphere model, more complicated factors have to be considered. One possibility is to introduce energy dissipation (e.g. proton-neutron collisions, internal shocks, or magnetic reconnections) and particle heating around the photosphere region. Such a dissipative photosphere model can naturally account for a high energy tail through Compton scattering, but could not significantly modify the low-energy spectral index from $\alpha \sim +0.5$. \cite{vurm11} introduced a synchrotron emission component, which peaks below the quasi-thermal component to make the ``effective'' low-energy spectral index softer. In order to make this synchrotron + quasi-thermal spectrum mimic a Band function as observed, the outflow magnetization parameter has to fall into a narrow range. Recently, \cite{thompson13} invoked a magnetically dominated, low baryon-loading outflow, and modified the low-energy spectral index through the contribution from electron-positron pairs. Several authors pointed out the contrived conditions for the dissipative photosphere models to produce a single-component spectrum \citep{vurm13,asano13,kumar13}. Another possibility to soften the spectrum below $E_p$ is to introduce a structured jet. \cite{lundman13} showed that $\alpha \sim -1$ can be reproduced given that the GRB jets have a near constant $L_w$ but a structured Lorentz factor profile with angle. This can enhance the high-latitude contribution (large $1/\Gamma$ cone at high-latitudes) to raise flux in large angles. For more general structured jets where both $L_w$ and $\Gamma$ follow a certain angular profile \citep[e.g.][]{meszaros98,zhangmeszaros02b,rossi02,zhang04}, the $\alpha$ value would not be very different from what is calculated in this paper. In all these models, it is unclear how the ``hard-to-soft'' $E_p$ evolution commonly observed in many GRB pulses can be accounted for.

Alternatively, the main Band-component in the GRB spectra could arise from an optically-thin region well above the photosphere due to synchrotron radiation. \cite{uhm13} recently showed that if the emission radius is large enough, the fast cooling problem for synchrotron radiation is alleviated, and $\alpha \sim -1$ can be reproduced in a moderately fast cooling regime. The hard-to-soft $E_p$ evolution pattern is a natural prediction in this model, since the outflow streams from small-radii where magnetic fields are stronger to large-radii where magnetic fields are weaker. Alternatively, the Band component may be interpreted as slow-cooling or slow-heating synchrotron emission in internal shocks where magnetic field strengths decays rapidly behind the shock \citep{peer06b,asano09,zhao14}.

Recent Fermi observations revealed a quasi-thermal component superposed on the main Band component in a growing population of GRBs \citep{ryde10,zhang11,guiriec11,axelsson12,guiriec13}. The spectral shape in our calculated photosphere emission is consistent with what is observed, suggesting that that component is very likely the photosphere emission from the GRB outflow \citep{peer12}. This component is typically weaker than what is predicted in the standard fireball-internal-shock model, so that a certain degree of magnetization is needed for the outflow \citep{zhangpeer09}. Within this picture, the non-thermal emission region in the optically-thin zone could be the internal shock region only if the magnetization parameter already falls below unity at the internal shock radius \citep{daigne11}. It is possible that the outflow is still moderately magnetically dominated in the large zone. In this case, efficient GRB emission is possible due to internal-collision-induced magnetic reconnection and turbulence (ICMART) \citep{zhangyan11,zz14}.

\medskip
We thank Xue-Feng Wu, Asaf Pe'er, Andrei M. Beloborodov, Z. Lucas Uhm, He Gao, Wei-Hua Lei, Hou-Jun L\"u, and Bin-Bin Zhang for helpful discussion or comments, and an anonymous referee for helpful suggestions. This work is partially supported by NASA under grant NNX10AD48G.

\appendix

Plugging in $P(\nu,T)$, $\hat P(r,\Omega)$ in Equation (\ref{eq:f_nu_2_3_1}), one gets
\begin{equation}
\begin{array}{cl}
\hat F_{\rm \nu}(\nu,t,\hat t) & =  {\dot N_0 \over 4 \pi d_{\rm L}^2} \int \int \hat P(r,\Omega)\cdot { n_\gamma(\nu,T) \over 16\pi({kT \over hc})^3\cdot \zeta(3)}\cdot h\nu   \cdot
\delta \left(t - \hat t - ({r u \over \beta c}-t_0) \right)d\Omega dr\\ \\
& ={\dot N_0 \over 4 \pi d_{\rm L}^2} \int_{\rm r_{min}}^{\rm r_{max}} \int \int_0^{\rm 2\pi} {\sigma_{\rm T} n' \Gamma {{\cal D}^2 \over 4\pi} e^{-\rm \tau(r,\mu,r_{\rm out})} \over A} \cdot { n_\gamma(\nu,T) \over 16\pi({kT \over hc})^3\cdot \zeta(3)}\cdot h\nu   \cdot
\delta \left(t - \hat t - ({r u \over \beta c}-t_0) \right)d(-\mu) d\phi dr\\ \\
& ={\dot N_0 \over 4 \pi d_{\rm L}^2} \int_{\rm r_{min}}^{\rm r_{max}} \int {\sigma_{\rm T} n' \Gamma {{\cal D}^2 \over 2} e^{-\rm \tau(r,\mu,r_{\rm out})} \over A} \cdot { n_\gamma(\nu,T) \over 16\pi({kT \over hc})^3\cdot \zeta(3)}\cdot h\nu   \cdot
\delta \left(t - \hat t - ({r u \over \beta c}-t_0) \right)d(-\mu) dr\\ \\
& ={\dot N_0 \over 4 \pi d_{\rm L}^2} \int_{\rm r_{min}}^{\rm r_{max}} \int {\sigma_{\rm T} n' \Gamma {1 \over 2 \beta (\Gamma u)^2} e^{-\rm \tau(r,\mu,r_{\rm out})} \over A} \cdot { {8 \pi {\nu}^2 \over c^3} {1 \over {\rm exp} ({h\nu \over kT} )-1} \over 16\pi({kT \over hc})^3\cdot \zeta(3)}\cdot h\nu   \cdot \delta \left(u - { \beta c (t - \hat t + t_0) \over r} \right){\beta c \over
r}du dr ~~~~(u=1-\beta \mu)\\ \\
& ={\dot N_0 \over 4 \pi d_{\rm L}^2} \int_{\rm r_{min}}^{\rm r_{max}} \int {\sigma_{\rm T} n' \Gamma {1 \over 2 \beta (\Gamma u)^2} e^{-\rm \tau(r,\mu,r_{\rm out})} \over A} \cdot { {8 \pi {\nu}^2 \over c^3} ( {\rm exp} ({h\nu \Gamma u \over kT^\prime} )-1)^{-1} \over 16\pi({kT^\prime \over hc \Gamma u})^3\cdot \zeta(3)}\cdot h\nu   \cdot \delta \left(u - { \beta c (t - \hat t + t_0) \over r} \right){\beta c \over
r}du dr ~~~~(T={T^\prime \over \Gamma u})\\ \\
& ={\dot N_0 \over 4 \pi d_{\rm L}^2} \int_{\rm r_{min}}^{\rm r_{max}} {\sigma_{\rm T} n' \Gamma {r^2 \over 2 \beta (\Gamma \beta c (t - \hat t + t_0))^2} e^{-\rm \tau(r,\mu,r_{\rm out})} \over A} \cdot { {2 {\nu}^2 \over c^3} ({ {\rm exp} ({h\nu \Gamma  \beta c (t - \hat t+t_0) \over kT^\prime r} )-1})^{-1} \over 4({kT^\prime r \over hc \Gamma \beta c (t - \hat t+t_0)})^3\cdot \zeta(3)}\cdot h\nu {\beta c \over
r} dr.\\ \\
\end{array}
\label{eq:f_nu2}
\end{equation}

The limits of integration can be calculated from the formula of equal arrival time surface, $t - \hat t = {r u \over \beta c}$, i.e.  $r={\beta c (t - \hat t) \over u}$. With $\theta_{\rm min}=0$ and $\theta_{\rm max}=\pi/2$, we get $r_{\rm min}={\rm max}[\beta c (t - \hat t),r_0]$ and $r_{\rm max}={\rm max}[{\beta c (t - \hat t) \over 1- \beta},r_0]={\rm max}[\Gamma^2 (1+\beta) \beta c (t - \hat t),r_0]$.

%\clearpage
%*******************************************************************************************************************
%*******************************************************************************************************************
%\clearpage

%\bibliographystyle{apj}

%\bibliography{ms}

%*******************************************

\end{CJK*}
\end{document}